\documentclass[10pt,final]{IEEEtran}
\author{
}
\title{TB-ICT: A Trustworthy Blockchain-Enabled System for Indoor COVID-19 Contact Tracing\thanks{This Project was partially supported by the Department of National Defence's Innovation for Defence Excellence and Security (IDEaS) program.}}
\author{Mohammad Salimibeni$^\dagger$, \textit{Graduate Member, IEEE}, Zohreh Hajiakhondi-Meybodi$^{\ddagger}$, \textit{Graduate Member, IEEE},  Arash Mohammadi$^\dagger$,  \textit{Senior Member, IEEE}, and Yingxu Wang$^{\dagger\dagger}$, \textit{Fellow,~IEEE}\\
$~^\dagger$Concordia Institute for Information Systems Engineering, Concordia University, Montreal, Canada\\
$~^{\ddagger}$ Department of Electrical and Computer Engineering, Concordia University, Montreal, Canada\\
$~^{\dagger\dagger}$ Department of Electrical and Computer Engineering, University of Calgary, Calgary, AB, Canada
}
\usepackage{epsfig}
\usepackage{amsmath}
\usepackage{multirow}
\usepackage{amsthm}
\usepackage{amssymb }
\usepackage{amsmath}
\usepackage{bm}
\usepackage{cite}
\usepackage{float}
\usepackage{mathrsfs}
\usepackage{amsfonts}
\usepackage{algorithm2e, algpseudocode}
\usepackage{subfigure}
\usepackage[usenames]{color}
\usepackage[dvipsnames]{xcolor}
\def\Bc{_{B_c}}
\def\uu{_{u}}

\def\ICT{\text{TB-ICT}}

\def\i{_{i}}
\def\j{_{j}}

\def\t{(t)}

\def\u{\bm{\mu}}

\def\t{(t)}

\begin{document}

\date{\today}
\maketitle
\thispagestyle{empty}

\begin{abstract}
Recently, as a consequence of the COVID-19 pandemic, dependence on Contact Tracing (CT) models has significantly increased to prevent spread of this highly contagious virus and be prepared for the potential future ones. Since the spreading probability of the novel coronavirus in indoor environments is much higher than that of the outdoors, there is an urgent and unmet quest to develop/design efficient, autonomous, trustworthy, and secure indoor CT solutions. Despite such an urgency, this field is still in its infancy. The paper addresses this gap and proposes the Trustworthy Blockchain-enabled system for Indoor  Contact Tracing ($\ICT$) framework. The $\ICT$ framework is proposed to protect privacy and integrity of the underlying CT data from unauthorized access. More specifically, it is a fully distributed and innovative blockchain platform exploiting the proposed dynamic Proof of Work (dPoW) credit-based consensus algorithm coupled with Randomized Hash Window (W-Hash) and dynamic Proof of Credit (dPoC) mechanisms to differentiate between honest and dishonest nodes. The $\ICT$ not only provides a decentralization in data replication but also quantifies the node's behavior based on its underlying credit-based mechanism. For achieving high localization performance, we capitalize on availability of Internet of Things (IoT) indoor localization infrastructures, and develop a data driven localization model based on Bluetooth Low Energy (BLE) sensor measurements. The simulation results show that the proposed $\ICT$ prevents the COVID-19 from spreading by implementation of a highly accurate contact tracing model while improving the users' privacy and security.
\end{abstract}

\section{Introduction} \label{sec:Intro}
Novel Coronavirus disease (COVID-19) has abruptly and undoubtedly changed the world as we knew at the end of the $2^{\text{nd}}$ decade of the $21^{\text{st}}$ century. Highly contagious nature of the novel coronavirus has resulted in a global outbreak and has significantly imperiled global healthcare systems~\cite{Wang2020-2, Mahmud2020}. On account of such a highly contagious disease, people around the world are being asked to practice strict social distancing protocols. In this regard, Contact Tracing (CT)~\cite{Ferretti2020, WorldWHO2020, Hellewell:Health2020}, i.e., contact identification, contact listing, and follow-ups, becomes of paramount importance. Manual CT solutions~\cite{Shubina2020} are labor-intensive, making it inefficient, error-prone, and time-consuming in addition to suffering from potential security and privacy issues. For these reasons, the focus of recent research works~\cite{Braithwaite2020} have been shifted towards development of autonomous CT models. Using technologies such as Internet of Things (IoT), mobile applications, and/or wearable devices, efficient autonomous CT solutions~\cite{Shubina2020, Scudellari2020} can be implemented to immediately inform potential users at risk, while reducing the required amount of labor force. Since the spreading probability of COVID-19 in indoor environments is higher than that of outdoors~\cite{Nishiura2020}, there is an urgent and unmet quest to develop/design trustworthy indoor CT solutions with highest achievable tracking capabilities. To achieve these objectives (i.e., trustworthiness and high indoor localization accuracy), we focus on (i) Blockchain-enabled design, coupled with (ii) Angle of Arrival (AoA)-based localization via Bluetooth Low Energy (BLE) beacons.

With regards to the Objective (i), a critical aspect for wide-speared adoption of indoor CT models is trustworthiness, i.e., assuring and preserving users' personal data~\cite{Song2020, Chan:MCT2020}. Several countries and related agencies~\cite{Cho2020} have released recent reports on security and privacy breaches of CT applications. Compliance with country-level health and privacy legislation, however, have yet to be discussed within CT platforms. Decentralized data storage, Decentralized Applications (DApps)~\cite{Scudellari2020, Shubina2020}, and Blockchain-based methods~\cite{Zhaofeng2020} are potential solutions to address these issues. Among these technologies, Blockchain-enabled models are favored since they inherently share users' data by employing a decentralized structure and cryptographical encryption that protects the privacy and security of the shared information. The trust provided based on a secure/distributed platform that keeps the users' identity and personal information anonymously, makes blockchain approaches an attractive solution to address the aforementioned concerns~\cite{Chamola:2020}. In blockchain-based services, information can be validated, data storage is immutable, and the data can be updated in a real-time fashion. More importantly, blockchain-based applications are, typically, independent of third parties. Since all nodes share the exact copy of the original data, which is hashed and chained in different blocks of the blockchain, it is impossible to manipulate data stored within different blocks of the system. In summary, in a blockchain-based CT infrastructure, system's consistency is not at risk and is widely resistant to tampering~\cite{Choo2020}.

With regards to the second identified objective (Objective (ii), i.e., equipping indoor CT solutions with accurate localization), while existing Global Positioning Systems (GPS) can be used to provide the required accurate localization in outdoor environments, tracking in indoors is a different and challenging task. For the design of autonomous indoor CT solutions, pre-installed IoT sensors~\cite{Zhu2020, Sadowski2020} can be leveraged to use existing infrastructures and allow widespread and timely adoption. To identify, track, and monitor users' location, which is the input of an autonomous CT model, several IoT-based indoor localization technologies such as BLE~\cite{Spachos2020, Hajiakhondi:SMC2020}, and Ultra Wide Band (UWB)~\cite{Yousefi2015} frameworks are available. Among such technologies, BLE-based tracking is of particular interest due to its ubiquitousness in most modern smart devices, low power consumption, and low cost. AoA-based localization~\cite{Hajiakhondi:Fusion2020} is one of the most efficient triangulation frameworks to measure users' location through the direction of the incident signal. BLE-based AoA estimation, however, was not possible until recently that direction-finding feature is introduced to the Bluetooth $5.1$ Core specification. Consequently, research works in this area are still in their infancy. This motivates the design of a BLE-based AoA estimation to increase the accuracy of the proposed indoor CT model. Despite recent researches performed on blockchain-enabled CT models, there is no integrated infrastructure coupling secure contact identification, accurate indoor localization, and follow-ups via a decentralized secure approach. The paper addresses this gap.

\vspace{.025in}
\noindent
\textbf{Contribution:} The main contribution of the paper is implementation of an innovative trustworthy, Blockchain-enabled, and Smart Indoor Contract Tracing ($\ICT$) framework by integration of blockchain and IoT technologies. The proposed blockchain-enabled $\ICT$ framework is shown in Fig.~\ref{Localiz}. In summary, the paper makes the following key contributions:
\begin{itemize}
\item The $\ICT$ framework is proposed to protect privacy and integrity of the underlying CT data from unauthorized access.  The $\ICT$ is an alternative blockchain platform exploiting the proposed dynamic Proof of Work (dPoW) credit-based consensus algorithm coupled with a Randomized Hash Window (W-Hash) and dynamic Proof of Credit (dPoC) mechanisms to differentiate between honest and dishonest nodes. Unlike the existing CT solutions~\cite{Bay:2020, Apple:2020, Levy:2020, Mozur:2020}, which are centralized and potentially vulnerable to information leakage, $\ICT$ is a fully distributed framework without any reliance on third parties. The $\ICT$ not only provides a decentralization in data replication but also quantifies the node's behavior based on its credit-based mechanisms. A Proof of Concept (PoC) version of the $\ICT$ framework implementation is made publicly accessible\footnote{https://github.com/MSBeni/SmartContactTracing\_Chained}. In this PoC, sample localization data is fed to the local database of distributed nodes and the proximity of users is calculated and saved in the database.
\item To boost the accuracy of the $\ICT$ model, we design an efficient data driven AoA-based indoor localization framework. More specifically, to avoid the need to analytically model noise, path-loss, and multi-path effects, a Convolutional Neural Network (CNN) model is designed on the subspace-based angle estimation in a 3-D indoor environment. For robust localization even in the worst-case scenario, we consider an indoor environment without presence of Line of Sight (LoS) links affected by Additive White Gaussian Noise (AWGN) with different Signal to Noise Ratios (SNRs). Moreover, the destructive effect of the elevation angle of the incident signal is considered.
\end{itemize}
The rest of the paper is organized as follows: In Section~\ref{sec:RWs} related works are discussed. Section~\ref{sec:UBFusion} formulates the BLE signal to extract location information and introduces the proposed CNN-based AoA localization framework. In Section~\ref{sec:BC}, we introduce the proposed blockchain-based $\ICT$ framework. Simulation results are presented in Section~\ref{sec:sim}. Finally, Section~\ref{sec:con} concludes the~paper.

\vspace{-.1in}
\section{Related Works} \label{sec:RWs}

Four of the most well-known CT applications recently proposed are TraceTogether from Singapore~\cite{Bay:2020}, Google/Apple joint project on a CT platform~\cite{Apple:2020}, COVID-19 application proposed by National Health Service (NHS)~\cite{Levy:2020}, and Chinese application named Health code system~\cite{Mozur:2020}. TraceTogether is a centralized service that uses Bluetooth technology to discover and store close proximity of users locally. BLE security concerns, e.g., bugging, sniffing, jamming, and power consumption in active advertisement fashions, are the issues related to this approach. Apple and Google exploit BLE technology as well. Unlike the TraceTogether, Apple/Google platform does not hold the user's real identity/data, but a central server is used for contact profiling and sending necessary notifications. The vulnerability in user information leakage, which could result in privacy issues, is also the same in the NHS COVID-19 application. On the other hand, the health code system leveraged a different approach in CT, utilizing QR code-based relational cross-match instead of BLE. While the health code is a better platform due to its lower power consumption, the centralization of the platform and violating the importance of anonymity of the users are key security issues of this platform. In what follows, first, we provide a brief overview of AoA and BLE-based localization, then we outline recent literature on blockchain-based approaches:

\vspace{.025in}
\noindent
\textbf{\textit{BLE Literature Review:}} High power consumption is one of the most important challenges of wireless technology for IoT applications. BLE technology has opened the way for IoT applications to flourish by enabling low-energy communications. BLE uses the same frequency spectrum as WiFi, ranging from $2.4$ to $2.48$ GHz, with $37$ data channels and $3$ advertisement channels, each with a bandwidth of $2$ MHz~\cite{Hajiakhondi:SMC2020}. There are several BLE-based indoor localization frameworks including but not limited to Received Signal Strength (RSS)-based methods~\cite{Mehryar:2019}, Time of Arrival (ToA)~\cite{Spachos:2018}, and Angle of Arrival (AoA)~\cite{Hajiakhondi:Fusion2020}. AoA-based localization is an active research field, where among all possible approaches, subspace-based angle estimation algorithms~\cite{Cheng:2015, Shafin:2016}, such as MUltiple Signal CLassification (MUSIC) and its extensions, are among the most reliable methods, especially in the presence of noise. Subspace-based algorithms, however, are prone to the multi-path effect, which is an unavoidable factor in indoor environments~\cite{Yassin}. Under the assumption that there is a strong LoS path between the transmitter and the receiver, there are a variety of approaches to mitigate the multi-path effect, including channel classification~\cite{He:2014}, Kalman filter-based techniques~\cite{Zhang:2016}, and subsample interpolation methods~\cite{Exel:2014}. There are, however, key challenges ahead: On the one hand, due to the computational complexity of the multi-path identification/mitigation methods, it is challenging to analytically model and compensate for the multi-path effect. On the other hand, ensuring that the receiver receives at least one LoS signal is a particular challenge. For these reasons, the focus has shifted to data-driven approaches such as those based on Artificial Intelligence (AI)~\cite{Wang2017,Wang2018}. Consequently, the need for complex and precise analytical models is reduced by considering the effects of multi-path, noise, and path-loss on the train dataset of the AI models. In this context, Reference~\cite{Comiter:2018} introduced a CNN-based AoA localization method in a 2-D indoor environment, where the transmitted signal is only affected by noise. Although the elevation angle of the incident signal is not used for localization purposes, it leads to a considerable location error. Therefore, it is essential to consider a 3-D indoor environment. For this reason, the effect of noise on the estimated angle in a 3-D indoor environment measured by DNNs was investigated by the authors in~\cite{Khan:2019}. Similarly, authors in~\cite{Wang2017, Wang2018, Hsieh:2019} proposed a DNN-based localization framework, where the Channel Impulse Response (CSI) is the input of the DNN, which is prone to noise,~shadowing, and small-scale fading, increasing the localization error.

\vspace{.025in}
\noindent
\textbf{\textit{Blockchain Literature Review:}} Blockchain is a distributed ledger that uses cryptography, Public Key Infrastructure (PKI), economic modeling, and a shared consensus mechanism to synchronize a distributed database. Blockchain can be considered as a Peer-to-Peer (P2P) network, operating on a large number of distributed devices to securely store/manage information, and transfer immutable append-only transaction logs, which are signed cryptographically~\cite{Nathan2019}. More precisely, blockchain is a hierarchical chain of blocks, where each block stores transactional records with transactions being distributed across the network to improve robustness against unauthorized modifications. Each block header consists of several fields, including the hash field (representing the block's hash value), a field to store hash value of the previous block, the nonce field for consensus algorithm, and the main body field. Hash-value is utilized to secure the constituent block. In this regard, each block appends the hash value of its preceding to provide a secure chain. To ensure integrity of transactions, blocks are included in the distributed ledger with time-stamp and hash value~\cite{Choo2020}. Each block needs to be mined and validated by all mining nodes across the network, after which the block is appended to the chain and cannot be changed or manipulated afterwards. This process maintains data integrity and provides a tamper-resistant network. Unique characteristics of blockchain models such as decentralized structure, fault-tolerance, persistency, anonymity, transparency, and immutability~\cite{Choo2020} make them the ideal technology for IoT-based applications~\cite{Javaid2020}, especially, for securely storing CT information in a trustworthy and distributed fashion without dependence on a central processing~unit.

Recently, blockchain-based and privacy-preserving approaches~\cite{Nguyen:2020} are proposed in the wake of the COVID-19 outbreak. The main objective is to protect the identity of infected users and prevent the release of fake information or false claims that can produce panic in public. Blockchain-based localization~\cite{Amoretti2018} and CT~\cite{Martinez2020, Song2020, Xu2020} have been discussed in recent literature. Most recent researches~\cite{Song2020, Xu2020}, however, mainly focused on data security and location privacy in smart CT services. For instance, Martinez \textit{et al.}~\cite{Martinez2020} used blockchain to store digitally signed pairwise encounters, which are transferred between users' handheld devices, and also leveraged AI approaches to recognize close contacts. Moreover, the platform is focused on the Sybil attack not deploying localization approaches and designing a privacy-preserving blockchain-based platform for CT purposes. Xu \textit{et al.}~\cite{Xu2020}, on the other hand, proposed the BeepTrace, which is a platform focusing on user's privacy in the whole CT system. Two blockchains for notification and tracing are implemented, and different positional technologies are used, yet no elaboration is made on the indoor localization approaches. The same issue is observed with Reference~\cite{Song2020}, where the authors represent a unified blockchain system for proximity-based CT and location-based CT considering Bluetooth technology. In~\cite{Micali2020}, a public blockchain is designed to enable each user to upload information, i.e., the hash of the encounters list, about the interaction with others at certain distances and for a certain period of time as qualified encounters. Avitabile \textit{et al.}~\cite{Avitabile2020} defined a robust CT proposal based on BLE leveraging the Diffie-Hellman (DH) secret sharing protocol to protect the system from more sophisticated attacks. Authors in~\cite{Lv2020}, proposed Bychain as a permissionless blockchain for tracing BLE-enable nodes or Long-Term Evolution (LTE) stations or WiFi access points, based on the location. Some certified nodes collect the information about the close interactions and submit them to the blockchain, and smart contracts will be used for the proof of location without  violation of  users' privacy. PriLok is proposed in~\cite{Esteves2020}, which is a permissioned blockchain that should be handled by collaboration of various entities, e.g., health and public authorities and telecommunication firms. PriLok uses a cellular network with the idea that the BLE technology and smartphones are not available as they should be specifically for older people. 

\section{The $\ICT$ Localization Model}\label{sec:UBFusion}
\begin{figure*}[t!]
\mbox{\subfigure[]{\includegraphics[scale=0.33]{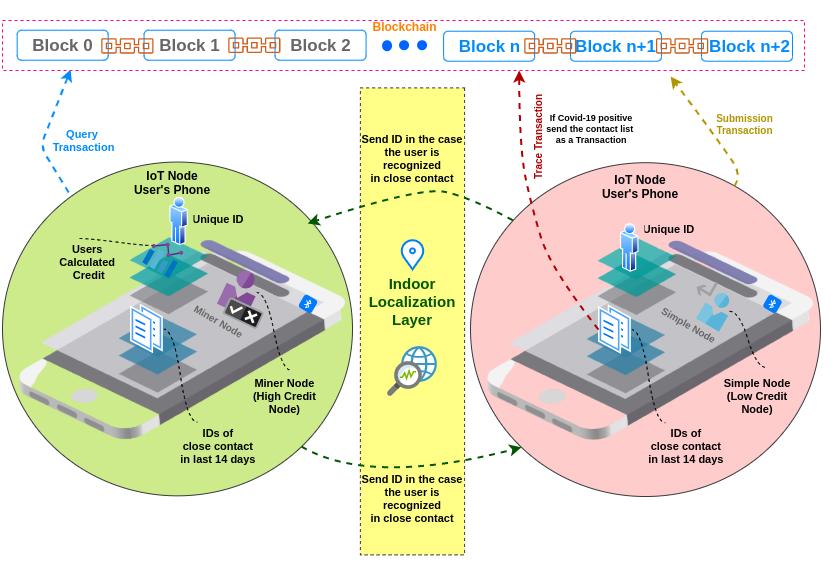}}
\subfigure[]{\includegraphics[scale=0.33]{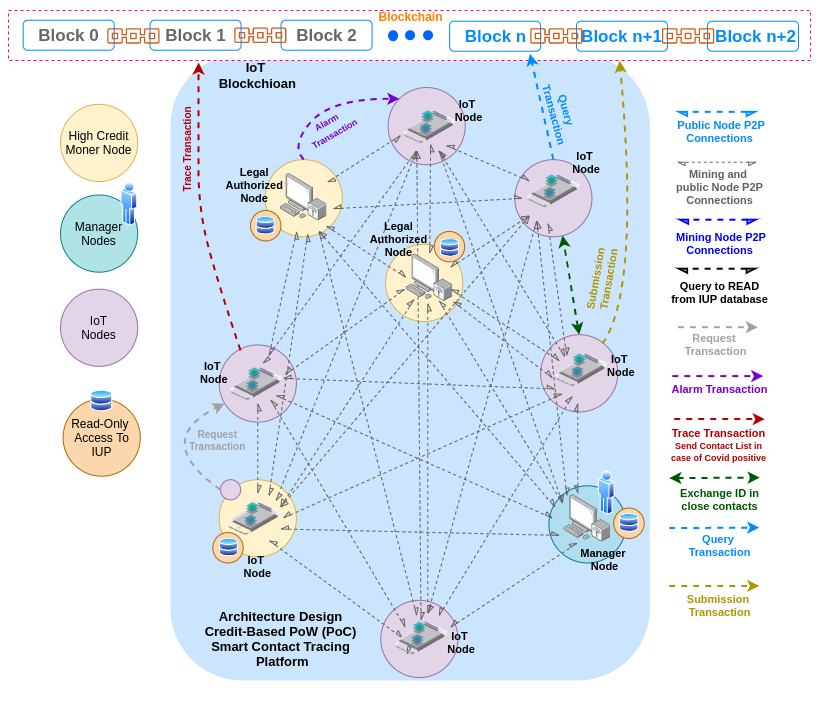}}}\\
\vspace{-.15in}
\caption{\footnotesize (a) $\ICT$ Indoor Localization Platform. (b)  $\ICT$ Blockchain network.}
\label{Localiz}
\end{figure*}
The structure of the $\ICT$ framework, as shown in Fig.~\ref{Localiz}, consists of the following two main components: (i) Indoor localization platform, and; (ii) The Blockchain network. In this section, we focus on the former category, i.e., the localization, where as can be seen from Fig.~\ref{Localiz}(a), the IoT nodes  (the BLE-enabled mobile devices)  exchange their unique ID with other nodes within their local neighborhood. To have an efficient indoor CT model, it is essential to implement a robust localization framework with high accuracy. The proposed localization solution is described below.

\subsection{CNN-based AoA Localization Framework}
\begin{figure*}[t!]
\centerline{\includegraphics [scale = 0.4] {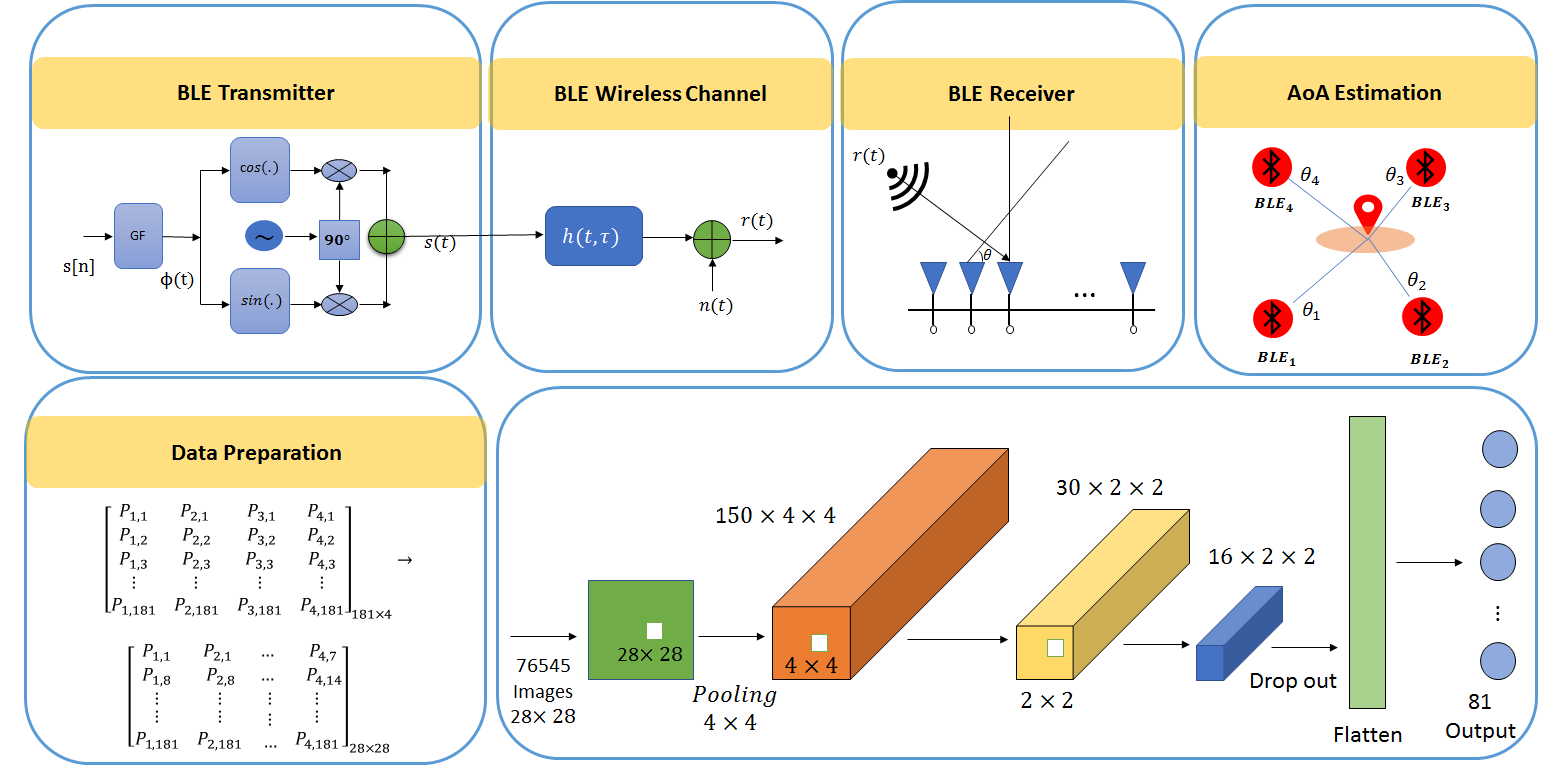}}
\vspace{-.1in}
\caption{\footnotesize Block diagram of the BLE transceiver, wireless channel model, and the proposed CNN-based AoA localization framework.}\label{BigPicture}
\vspace{-.2in}
\end{figure*}

In this subsection, first we formulate BLE wireless signal model. Then, we present our proposed CNN-based AoA localization framework relying on BLE technology. Fig.~\ref{BigPicture} illustrates different components of the proposed localization platform, i.e., (1) BLE transmitter; (2) BLE Wireless channel; (3) BLE receiver; (4) AoA estimation; (5) Data Preparation, and; (6) CNN-based localization, which are introduced below:

\vspace{.025in}
\noindent
\textbf{\textit{1. BLE Transmitter:}}
The baseband version of the transmitted signal $s_b(t)$ is calculated as
\begin{eqnarray}\label{e5}
\lefteqn{s_b(t) =s_i^b(t)+j s_q^b(t)\nonumber}\\
&=&\sqrt{\dfrac{2E}{T}} \Big\{\cos(\phi(t) + \phi_0)+j \sin(\phi(t) + \phi_0)\Big\}.
\end{eqnarray}
where terms $E$ and $T$ represent the energy and the time interval of the transmitted symbol, respectively. Term $\phi_0$ is the initial phase of the transmitted signal. Finally, term $\phi(t)$, denoting the phase deviation, is expressed as
\begin{equation}\label{e2}
\phi(t)=\dfrac{\pi h}{T} \int_{-\infty}^{t} \sum_{n=-\infty}^{+\infty} s[n] p(\tau - nT) d\tau,
\end{equation}
where $h$, known as the modulation index, is between $0.45$ to $0.55$ in BLE standard. Terms $s[n]=\pm 1$ and $p(t)$ denote the baseband pulse sequence and Gaussian Filter (GF), respectively. Finally, the transmitted BLE signal $s(t)= Re\{ s_b(t)e^{j2\pi f_c t}\}$, is  modulated by Gaussian Frequency-Shift Keying (GFSK), i.e.,
\begin{equation}\label{e1}
s(t)=\sqrt{\dfrac{2E}{T}}\cos \left(2\pi f_c t+ \phi(t) + \phi_0 \right),
\end{equation}
where $2.4 \leq f_c \leq 2.48$ GHz denotes the carrier frequency.

\vspace{.025in}
\noindent
\textbf{\textit{2. BLE Wireless Channel:}}
Because of the obstacles in indoor environments, a number of phase delayed and power attenuated versions of the transmitted BLE signal, which are affected by AWGN, are received by the BLE beacons. Therefore, the received BLE signal can be expressed as
\begin{equation}\label{received}
r(t)=\sum_{k=1}^{N(t)} \rho_k(t,\tau) s(t-\tau_k(t))+n(t),
\end{equation}
where $N(t)$ represents the number of detachable paths. Terms $\rho_k(t,\tau)$ and $\tau_k(t)$ denote the attenuation and the delay of the $k^{\text{th}}$ path, respectively. In addition, term $n(t)$ represents the AWGN channel, which is modelled by $n(t) \sim \mathcal{N}(0,\sigma ^2)$. In a BLE positioning system, the location of users can be obtained from the estimated CIR, i.e.,
\begin{equation}\label{e8}
h(t,\tau)=\sum_{k=1}^{N(t)}\rho_k(t,\tau) \delta(t-\tau_k(t)).
\end{equation}

\vspace{.025in}
\noindent
\textbf{\textit{3. BLE Receiver:}}
To extract the angle of the incident signal, BLE beacons are required to be equipped with an antenna array, where commonly the Linear Antenna Array (LAA) is used. In a typical LAA, there are $N_e$ number of elements with the same distance $d_0$, where the transmitted BLE signal is received by distinct elements with different phase differences. The discrete received signal by element $e$, sampled at time slot $m$, denoted by $r_e[m]$, is obtained as
\begin{equation}\label{e10}
r_e[m]=s^{'}[m] \Theta(\theta,\phi)[m] +n[m],
\end{equation}
where $\mathbf{\Theta}(\theta,\phi)$ denotes the array vector, expressed as follows
\begin{eqnarray}
\mathbf{\Theta}(\theta,\phi)= \exp([-j \dfrac{2\pi d}{\lambda}\cos \theta \cos \phi, \\ \nonumber
-j \dfrac{2\pi d}{\lambda}\sin \theta \cos \phi, -j \dfrac{2\pi d}{\lambda} \sin \phi]^T),
\end{eqnarray}
where $\theta$ and $\phi$ represent the azimuth and elevation angles, respectively. Term $d$ indicates the space between two consecutive elements of the LAA, which is equal to $\frac{\lambda}{2}$, with $\lambda=c/f_c$. Finally, term $c = 3 \times 10^8$ $m/s$ is the speed of light. By assuming $M$ samples in each received signal, we have
\begin{eqnarray}
\textbf{r} &=& [r_1[m] \ldots r_{N_e}[m] ]^T, \label{e6}\\
\text{and }\quad \textbf{s}^{'} &=& [s^{'}_1[m] \ldots s^{'}_{N_e}[m]]^T. \label{e7}
\end{eqnarray}
Therefore,  the received signal can be expressed as follows
\begin{equation}\label{e5}
\textbf{r}=\mathbf{\Theta}(\theta,\phi) \textbf{s}^{'}+\textbf{n}.
\end{equation}

\vspace{.025in}
\noindent
\textbf{\textit{4. AoA Estimation:}}
Given the angle of the incident signal through a subspace-based angle estimation algorithm, we propose to use a CNN model for localization. In this regard, the spatial spectrum of the received signal, denoted by $\textbf{p}(\theta,\phi,t)$, is calculated in each time slot as follows
\begin{equation}\label{e11}
\textbf{p}(\theta,\phi,t)=\dfrac{1}{\mathbf{\Theta}^H(\theta,\phi,t)\textbf{E}_N \textbf{E}_N^H \mathbf{\Theta}(\theta,\phi,t)},
\end{equation}
where $\textbf{E}_N$ is the noise eigenvectors of the covariance matrix $\textbf{R}=E[\textbf{r},\textbf{r}^H]$, and $\textbf{r}$ denotes the received signal.

\vspace{.025in}
\noindent
\textbf{\textit{5. Data Preparation:}}
%
In each time slot, a reshaped angle image is the input of the CNN, which is generated by $\textbf{P}(\theta,\phi,t)=[\textbf{p}_1(\theta,\phi,t), \ldots , \textbf{p}_{N_b}(\theta,\phi,t)]$, where $N_b$ indicates the number of available BLEs in the user's vicinity. Since $0 \leq \theta \leq 180$, there are $181$ samples in $\textbf{p}_i(\theta,\phi,t)$, where the minimum value of $\textbf{p}_i(\theta,\phi,t)$ indicates the incident angle $\theta$. Fig.~\ref{BigPicture}(b) illustrates a typical angle image, containing $724$ sets of AoA measurements generated by the subspace-based algorithm. As can be seen in Fig.~\ref{BigPicture}(b), the original angle image is of size $4 \times 181$, reshaped to be an square angle image of size $28 \times 28$ by zero padding.
This completes presentation of the localization platform of the $\ICT$. Next, we focus on the blockchain module.

\section{The $\ICT$ Blockchain Platform}\label{sec:BC}
The $\ICT$ network is a blockchain infrastructure where each entity act as a node in the system. The nodes are divided into light nodes and full ones, which is a regular categorization within the blockchain context. Several nodes participating in the network are power-constrained mobile devices and cannot store the whole network data as is the case in the classical blockchain frameworks. Referred to as light nodes (as shown in Fig.~\ref{Localiz}(a)), power constraint devices only store the data associated with the last $14$ days (the period defined and necessary for checking the infection) and can send and receive different transactions in the network, including but not limited to Submission Transaction (ST), Trace Transaction (TT), Query Transaction (QT), Request Transaction (RT), and Alarm Transaction (AT). Light nodes can also act as miners and run the consensus algorithm to mine the block, sign and broadcast it. The $\ICT$ network architecture is shown in Fig.~\ref{Localiz}(b), which consists of three main components:

\noindent
\textbf{\textit{(i) Mobile Devises:}} Users in the network acting as light nodes and commute in the environment. The proximity of these nodes to neighboring devices is estimated in a real-time fashion leveraging the $\ICT$ localization platform as discussed in Section~\ref{sec:UBFusion}. Their contacts will be preserved in their local device to be used, if necessary, in future transactions. These nodes will receive credits (positive or negative) over time based on their proximity-based localization results and/or their behavior in the network. Each mobile device creates a blockchain account and receives public and private keys ($P_k, S_k$) within the initialization phase as the node's unique identifier. The key/pair is used to sign the transactions and create a unique public identifier to be placed in the BLE advertisement packets and to be shared with its neighbours.

\noindent
\textbf {\textit{(ii) Authorized Static Nodes:}} Static nodes, i.e., the full nodes, are mostly the legally authorized units, which have read access to data in the Infected Users Pool (IUP). The infected persons' data, which is saved in the IUP, is distributed among these nodes and cannot be changed. When an authorized node receives data about the infection of a node (a user is recognized to be infected with COVID-19 in a test center), this data will be broadcasted to the other authorized nodes. Any claim for infection by the users in the network will be validated with these nodes, and in case of any false claim, the claiming node will be punished by receiving negative credit points.

\noindent
\textbf {\textit{(iii) Manager Static Nodes,}} are specific full nodes responsible for managing (add/delete) the authorized nodes in the network whose public key is hard-coded into the software. Manager node $M$ (denoted by $AN_M$) can lunch a transaction $TX$ to keep the record of the authorized nodes as follows
\begin{equation}\label{eb1}
TX = Sign_{SK_{AN_M}}\big(PK_{AN_1}, PK_{AN_2}, \ldots, PK_{AN_n}\big),
\end{equation}
where $SK_{AN_M}$ denotes the secret key of the manager and ($PK_{AN_1}, PK_{AN_2}, \ldots, PK_{AN_n}$) represents public key of the other authorized nodes. Since the manager uses his secret key, which cannot be forged to sign the transaction, other authorized nodes can simply fetch the list of the authorized nodes in the blockchain network published by the manager. The number of the manager nodes can be defined based on the distribution of the network and national/provincial policies.

\begin{figure}[t!]
\centerline{\includegraphics [scale = 0.35] {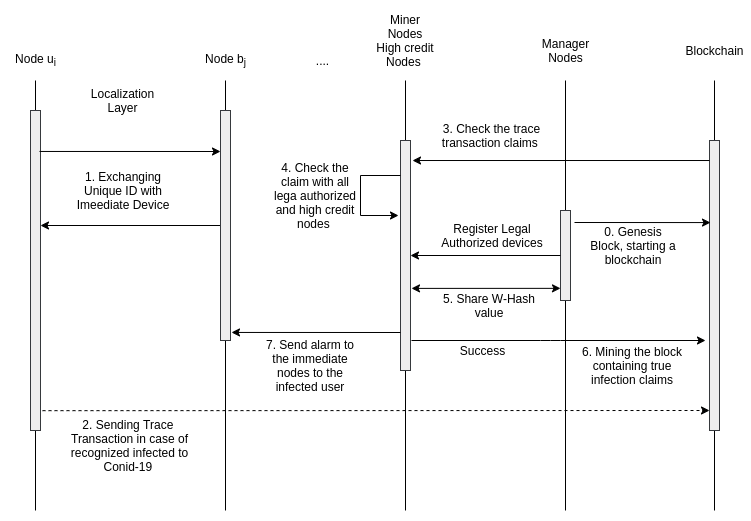}}
\vspace{-.15in}
\caption{\footnotesize Interconnections within the $\ICT$ network.}
\label{ICT-NI}
\end{figure}
Light nodes with high credit and authorized static nodes act as miners in the network. Other nodes can submit different transactions to know about the status of the network (e.g., QT and ST) and/or claim an infectious status as a transaction (i.e., the TT). Interactions between different nodes are shown in Fig.~\ref{ICT-NI}. The $\ICT$ architecture is distributed and resilient to potential cyber attacks. Attacks, such as DDoS, Sybil, or double-spending, however, would not be of interest for the malicious nodes, as the proposed platform is inherently designed to securely share and distribute data. There is no monetized structure to urge attackers targeting the system. In any case, the data itself is precious, and the users' information must be kept safe and secure.
In this context, most of the existing blockchain-based platforms use the Proof of Work (PoW) consensus protocol, which faces the following key~challenges:
\begin{itemize}
\item[(i)] \textit{Network Scalability:} The number of blocks that can be processed per unit time represents the throughput of the blockchain network, which shows the platform's scalability;
\item[(ii)] \textit{Security Vulnerabilities:} Keeping the submitted records and mined transactions immutable is a security-wise concern that should be proved/improved in a blockchain platform, and;
\item[(iii)] \textit{Computational Complexities:} The difficulty level determines the processing power (computational complexity) required to mine a new block.
\end{itemize}
To address these challenges, the proposed $\ICT$ is an alternative blockchain platform exploiting the Dynamic PoW (dPoW) Credit-based consensus algorithm along with a Randomized Hash Window (W-Hash), and Dynamic Proof of Credit (dPoC). Next, each of these components are detailed.

\subsection{Randomized Hash Window} \label{W-Hash}
\begin{figure}[t!]
\centerline{\includegraphics [scale = 0.3] {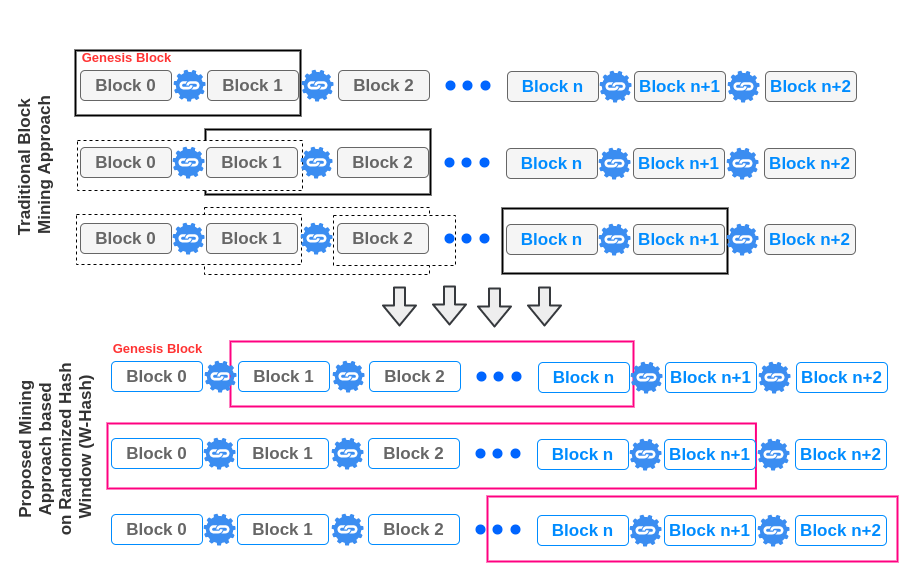}}
\vspace{-.15in}
\caption{\footnotesize $\ICT$ randomized hash window (W-Hash) solution to enhance the security of the proposed Blockchain network.\label{w-hash}}
\vspace{-.15in}
\end{figure}
The introduced Randomized Hash Window, referred to as the W-Hash, is a mechanism to mine a block, which is similar in nature to the randomized sliding window algorithm and the check-points concept~\cite{Javaid2020}. More specifically, a miner continuously mines the current block with a number, i.e., the nonce value, to meet the predefined Difficulty Level ($DL$) and produces the target hash string. W-Hash makes this hashing process harder, i.e., the W-Hash mechanism slides through the blockchain ledger similar to the sliding window algorithm. In other words, as can be seen in Fig.~\ref{w-hash}, the miner should hash the current block along with the sliding window of the previous blocks. For instance, assume W-Hash is $14$, and the current block number is $100$, the miner should consider the concatenation of block $100$ with previous $13$ blocks (from $87$ to $99$) to create the window size of $14$. Within the initializing phase of the $\ICT$ network, when a W-Hash is larger than the current number of blocks in the blockchain, the W-Hash will be set to zero. Otherwise, the current block will be hashed based on the previous succeeding blocks determined and concatenated considering the W-Hash. In the long term, when more blocks become added to the blockchain, the W-Hash, which is a value between $0$ to $100$, can be applied to every block. The drawback of such an approach is that the overhead complexity is of $O(n)$, where $n$ denotes the W-Hash value, when compared to the classical blockchain platforms with a constant PoW difficulty level. To address this issue, we propose dPoW difficulty managed by a credit-based evaluation mechanism as discussed next.

\subsection{Dynamic Proof of Work (dPoW)} \label{DPoW}
As stated previously, there is a tradeoff between security and scalability in blockchain networks. The proposed W-Hash solution is an innovative approach to maximize the security of the network. However, W-Hash directly affects the underlying computational complexity, mining costs, and mining time. As stated in~\cite{Javaid2020}, a dynamic approach can be employed to define the different levels to control the rate of the incoming communication traffic and the mining difficulty. Therefore, Dynamic PoW consensus approach and credit-based difficulty level ($DL$) determination for the nodes in the network is a solution to the added complexity of the W-Hash. Intuitively speaking, the users in the network with higher credit can exploit a much easier difficulty level in the system helping them to easily mine the blocks, and the confidential W-Hash value $n$ will be shared with them securely. On the other hand, a malicious miner now requires both $n-1$ blocks and the W-Hash value $n$ to mine a new block. A fraud process and/or faulty behaviors are, therefore, costly as they result in high negative credits for malicious nodes resulting in the higher difficulty.

In the $\ICT$ framework, we consider two difficulty levels to be applied for the nodes in different circumstances. As can be seen in Table~\ref{Table:table1}, Difficulty level $1$ ($DL_e$) has the lowest difficulty achieved by considering the first significant digit of the block hash string, which is calculated by incrementing the nonce continuously and check the results until meeting the difficulty constraints. The first level of difficulty $DL_e$ is assigned to the legally authorized nodes and the nodes in the network with a credit larger than a predefined threshold ($\alpha_d$). Details of the credit evaluation process will be covered later in Section~\ref{DPoC}. Due to the low level of difficulty, even devices with limited computational power (if they meet the constraint set by the credit threshold) can participate in the mining process. The second difficulty level ($DL_h$) is four times harder than $DL_e$, which is achieved by considering the first four significant digits of the target hash string. Level $4$ is considered for the low incoming communication traffic and for dishonest nodes with low credits.

\begin{table}[t!]
\centering
\caption{\small Difficulty Level (DL) of the Proposed Dynamic PoW mechanism.}\label{Table:table1}
\begin{tabular}{| l | l | l | l |}
\hline
\textbf{DL} & \textbf{Size (bits)} & \textbf{Target Hash} & \textbf{Difficulty} \\ \hline
$DL_e$ & 4 & SHA256[0:1] & High \\ \hline
$DL_h$ & 16 & SHA256[0:4] & Low \\ \hline
\end{tabular}
\vspace{-.2in}
\end{table}
\textit{Parameters of the dPoW}: Similar in nature to Bitcoin, data encryption is implemented in the $\ICT$ exploiting Elliptic Curve Cryptography (ECC) with Elliptic Curve Digital Signature Algorithm (ECDSA). The block size in the blockchain of Bitcoin is restricted to $1$ megabyte. Unlike the conventional blockchain model~\cite{Bitcoin, Zheng}, which has a fixed block size, $\ICT$ is designed to have a variable block size considering incorporation of the dPoW approach in its consensus algorithm. Given the variable block size of $\ICT$, we set an upper bound of $1$ megabyte for the block size. The variable block size is formulated as follows
\begin{eqnarray}\label{eb2}
\lefteqn{S\Bc = Overhead\Bc +\nonumber}\\
&& [(Data + EncryptionOverhead)\times N_{data}],
\end{eqnarray}
where $S\Bc$ is the current block size, $Overhead\Bc$ is the block metadata (number of bytes representing a block), $EncryptionOverhead$ is the number of bytes for current block data encryption, and $N_{data}$ is the number of data samples added to the current block. The difficulty level is defined as follows
\begin{equation}\label{eb3}
D\Bc = \frac{DL}{TargetHash},
\end{equation}
where $D\Bc$ represents the difficulty level of the current block, $DL$ is the level of the difficulty in the dPoW, i.e., $DL_e$ or $DL_h$, and the $TargetHash$ is the required string to mine a block. The mining interval of the blocks, representing the average time required to mine a block, can be formulated as
\begin{equation}\label{eb4}
Intervel_B = \frac{D\Bc \times 2^b}{HashRate},
\end{equation}
where $HashRate$ is a miner's computation power for hash iteration generation per second, and $ 2^b$ is responsible for difficulty level ($DL$) bits control, i.e., $16$ bits for the $DL_h$ and $4$ bits for $DL_e$. Table~\ref{Table:table1}, represents the difficulty level in the proposed dynamic structure.

\subsection{Dynamic Proof of Credit (dPoC)} \label{DPoC}
Consider a CT network with $N_{nodes}$ number of nodes. Node $i$, for ($1 \leq i \leq N_{nodes}$), is assigned a credit $CR\i\t$ at time instant $t > 0$, which is a parameter altering in real-time based on the node's interactions and behavior in the environment. Normal Behaviors, i.e., following the physical distancing regulations and sending transactions after being recognized as an infected person, will gradually increase the user's credit value over time. On the other hand, abnormal behaviors will diminish the node's credit over time. Abnormal behaviors can be classified into two main categories:
\begin {itemize}
\item \textit{Disease-Related Violations:} Breaching the physical distancing regulations and other related rules, which are mostly bond to the proximity-based localization results.
\item \textit{Anomalous Behavior in Blockchain Network:} Different anomaly behaviors in the networks including attacks or wrong claims, e.g., false claims of infection to the COVID-19 or failing to share contact information in the case of being diagnosed with COVID-19.
\end {itemize}
Node $i$, based on its behavior, can receive negative or positive scores resulting in the following credit
\begin{equation}\label{e1}
Cr\i\t= Cr^{Prox}\i\t + Cr^{-}\i\t,
\end{equation}
where $t$ is the current time index, $Cr^{Prox}\i\t$ is the proximity-based credit (which can be positive or negative), and $Cr^{-}\i\t \leq 0$ is the received negative credit score, which is calculated based on malicious behaviors of the node in the network. Following Section~\ref{sec:UBFusion}, let $\mathcal{O}\i\t = \{o_1\t, o_2\t, \ldots, o_{N\i\t}\t\}$, denote the set of $N\i\t$ BLE-enabled devices observed by the $i^{\text{th}}$ user (excluding the user itself) at time $t$. Every few seconds, this set of devices can be changed as the user moves within the environment. Based on the distance between user $i$ and a neighboring BLE-enabled device, Node $o\j$, ($1\leq j \leq N\i\t$), we consider two categories, i.e., immediate nodes (less than $2$ meters) and non-immediate node (near or far, when the distance is more than $2$ meters). We define $P\i(o\j)$ as the credit assigned to Node $i$ based on its proximity to the neighbouring Node $o\j \in \mathcal{O}\i\t$~as follows
\begin{eqnarray}\label{e2}
P\i(o\j\t) = \begin{cases} \frac{-1*\lambda^-}{L(i, o\j\t)} , & \text{Immediate}, \text{ }L(i, o\j\t)<2 \\ \frac{L(i, o\j\t)}{\lambda^+}, & \text{Near/Far}, \text{ }L(i, o\j\t)\geq2 \\\end{cases},
\end{eqnarray}
where $L(i, o\j\t)$ denotes the distance between Agent $i$ and its contact $o\j\t$ at time $t$. Terms $\lambda^-$ and $\lambda^+$ are the credit calculation coefficients that are defined based on the policy utilized to punish and reward the users. In a difficult situation when the disease is highly contagious and preventing policy should be considered more seriously, both $\lambda^-$ and $\lambda^+$ can have a higher value. As can be seen, keeping a healthy distance will be rewarded by a positive credit. On the other hand, in case of immediate proximity, the node will be charged with a higher negative credit score proportional to its closeness to the other contact. These coefficients allow the model to manage credit weight distribution. Considering Eq.~\eqref{e2}, the proximity credit assigned to Node $i$ is calculated as follows
\begin{equation}\label{e3-2}
Cr^{Prox}\i(t) = \sum_{o_j\t \in \mathcal{O}\i\t}P\i(o\j\t),
\end{equation}
where $Cr^{Prox}\i(t)$ is adjusted based on the level of healthy preventive behavior practiced by Node $i$. To measure the total credit value of a node based on the localization result, this value will be added to the previously calculated proximity-based credit to calculate the final credit. Based on the scanning and advertising intervals of the BLE devices, every few seconds, the new credit value of the user will be updated and added to the previously calculated proximity-based credit.

Disease-related violations are incorporated into the credit score based on the proximity results obtained from the localization model. Term $Cr^{-}_i\t$ is negatively proportional to the number of anomalous behaviors of Node $i$, to account for anomalous behaviors, i.e.,
\begin{equation}\label{e4}
Cr^{-}\i\t = -\sum_{l=1}^{m\i}\omega.\frac{\Delta T}{t-t_l}
\end{equation}
where $m\i$ is the total number of abnormal behaviors performed by user $i$, $t_l$ indicates time point of the $l^{\text{th}}$ abnormal behavior, and $\Delta T$ represents a unit of time. Term $\omega(x)$ is the penalty coefficient and is computed as follows
\begin{eqnarray}\label{e5}
\omega = \begin{cases} \omega_{FC} , & \text{If a false claim is made}\\ \omega_{SC}, & \text{If a contact violation happened}\\ \omega_{NA}, & \text{If a network attack happened}\end{cases},
\end{eqnarray}
where all coefficients ($\omega_{FC}$, $\omega_{SC}$ and $\omega_{NA}$) can be adapted based on the network's punishment policy. A node's credit can be measured via Eq.~\eqref{e1} by calculating $Cr^{-}\i\t$ and $Cr^{Prox}\i\t$. In the case of abnormal claims or malicious behavior of a node, the absolute value of $Cr^{-}\i\t$ becomes a large value, which will make continuous attacking impossible for a faulty node. The difficulty for a specific node is then set as follows
\begin{eqnarray}\label{e6}
D\uu\t = \begin{cases} DL1 , & \text{if } Cr\i\t \geq \alpha_d \\ DL2 , & \text{if } Cr\i\t < \alpha_d\end{cases}.
\end{eqnarray}
A new block is designed to be concatenated with $n_{wh}$ number of former blocks through the dPoW consensus replication approach. Let $B_j$ represent the $j^{\text{th}}$ block of the blockchain, $B_c$ the current block that should be mined, and $N_c$ as its nonce value. Hash of the current block ($B_c$) is formulated as~follows
\begin{eqnarray}\label{e7}
H_c = h\left(\sum_{j=1}^{n_{wh}-1}B_{c-j}+[B_c + N_c]\right),
\end{eqnarray}
where $H_{c}$ is the hash of the current block and $\sum_{j=1}^{n-1}B_{c-j}$ denotes the $n-1$ W-Hash blocks. If $H_c$ meets the difficulty level requirement, i.e., prefix zero length, the block can be mined, and the nonce value is successfully found. Adjusting the difficulty level based on the prefix zero is already discussed in Section~\eqref{DPoW}. A larger minimum required prefix zero makes it more difficult to find the nonce value and to mine a block. Therefore, honest nodes can consume less power in the mining procedure based on the proposed dynamic difficulty level mechanism. In contrast, the malicious nodes will be forced to consume more computational resources, making them unable to conduct any other abnormal behavior. Algorithm~\ref{Alg1} summarizes the block mining procedure of the $\ICT$ framework.

\begin{algorithm}[t!]
\SetAlgoLined
\KwData{$\ICT$ Block Mining Algorithm} \label{Alg1}
\KwResult{Target Hash String}
initialization\;
\textbf{Check Eligibility of the Node:}\\
\eIf{$Cr\u \geq \alpha_d$}{
\quad High Credit Node\;
\quad set target, $T = "0"$\;
\quad }{
\quad Low Credit Node\;
\quad set target, $T = "0000"$\;
\quad }
set nonce: $N_c = 0$\;
\While{mining}{
\For{block.index $\in [0,100]$}{
$H_c = h\left(\sum_{j=1}^{n_{wh}-1}B_{c-j}+[B_c + N_c]\right)$\;
\eIf{$H_c == SHA256[T]$}{
break\;
}{
$N_c++$\;
}
}
}
\caption{$\ICT$ Block Mining Algorithm}
\end{algorithm}

\section{Experiments}  \label{sec:sim}
\begin{figure*}[t]
\centerline{\includegraphics [scale = 0.38] {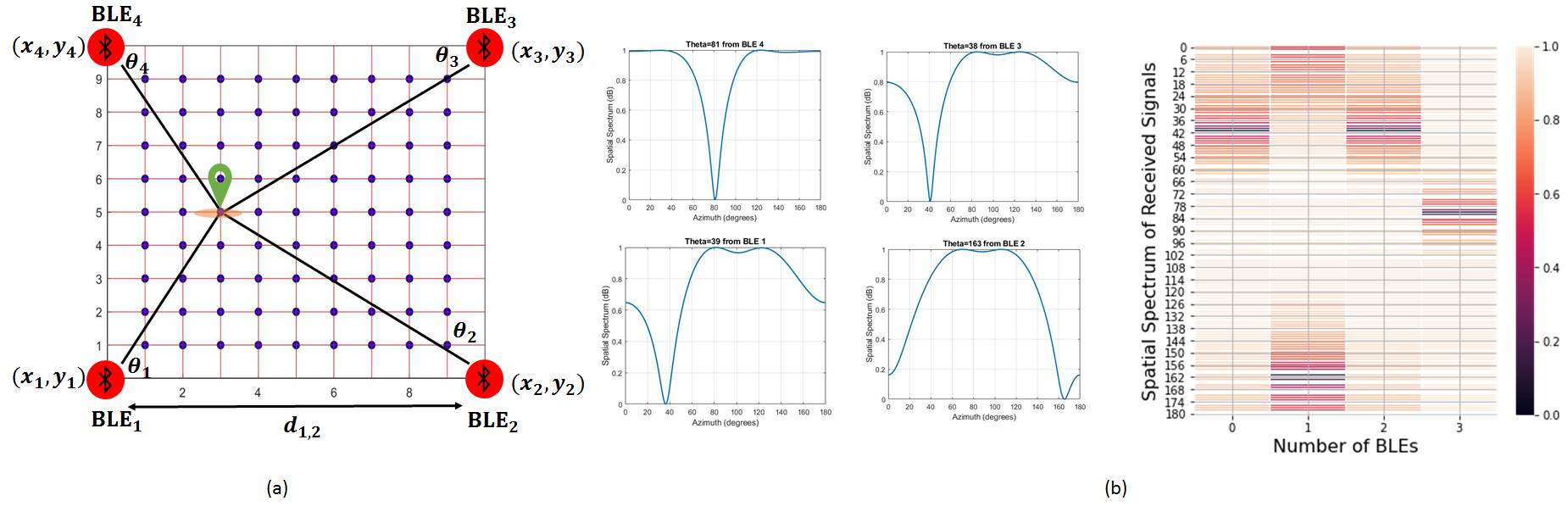}}
\vspace{-.15in}
\caption{\footnotesize (a) Experimental data collection of the CNN-based AoA localization framework. (b) An angle image, used as the input of the CNN-based framework.}\label{Angle}
\vspace{-.15in}
\end{figure*}
To evaluate the proposed $\ICT$ framework, we considered a real experimental testbed in a rectangular indoor area ($10 \times 10$) $m^2 $, divided into $400$ square zones each with dimension of ($0.5 \times 0.5$) $m^2 $ (see Fig.~\ref{Angle}(a)). There are $16$ BLE beacons, where the distance between each BLE beacon is equal to $4$ m. There exists $1,000$ number of users, where their movements are modeled by a random walk.

\begin{figure}[t!]
\centering
\mbox{\subfigure[]{\includegraphics[scale=0.33]{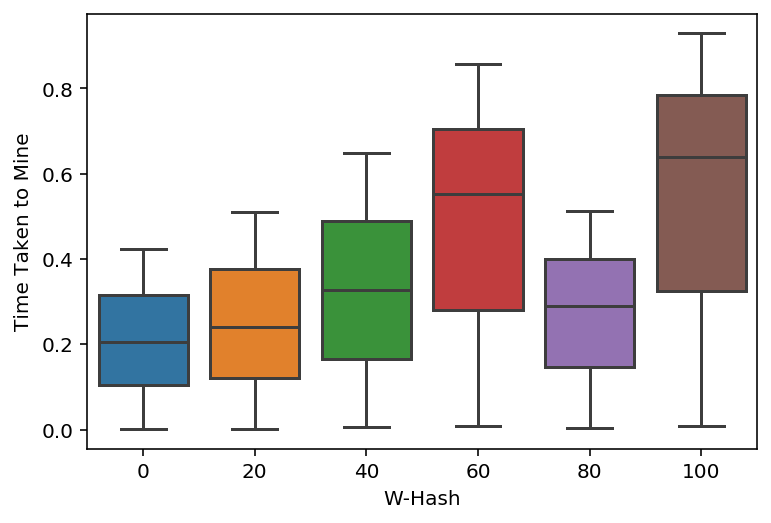}}
\subfigure[]{\includegraphics[scale=0.33]{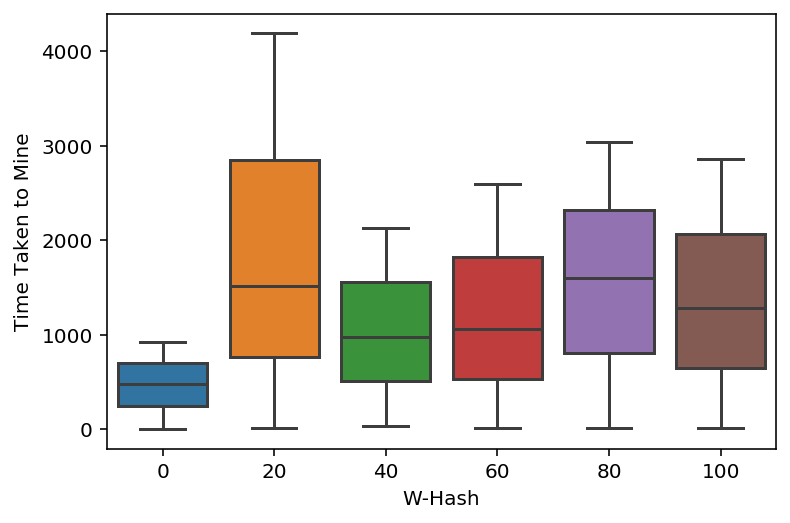}}}\\
\caption{\footnotesize Exploiting DPoW for different W-Hash values to mine $120$ sample blocks: (a) Difficulty Level $DL_e$ employed for high-credit and legally authorized nodes. (b) Difficulty Level $DL_h$ applied for low credit nodes (e.g., Malicious Nodes).}\label{Fig:W-Hash04}
\end{figure}
Six different W-Hash values ($0$, $20$, $40$, $60$, $80$, and $100$) are considered to evaluate the proposed system for its two different difficulty levels, i.e., $DL_e$ and $DL_h$ as are described in Table~\ref{Table:table1}. Fig.~\ref{Fig:W-Hash04} shows the different times spent to mine a block over different W-Hash values.  The W-Hash $0$, represents the classical PoW process while the other W-Hash values, concatenate the $n_{WH}$ number of blocks with the current block in order to enter the mining process. These plots are formed based on localization data of $1,000$ indoor users where all the nodes are active (have movements in the environment) and have several interactions. All the interactions are considered to be collected and used to calculate the users' credit over time exploiting Algorithm~\ref{Alg1}. All the proximities affect the credit value of the user where the close interactions will push negative scores to the credit, while near or far proximities (distances between $2$ to $10$ meters) will bring positive points to the user's credit. However, when a user is recognized to be infected with COVID-19, only the contacts in immediate distance to that user will be reported in a transaction to the blockchain. This will keep the block's incoming data as little as possible and improves the system's throughput.

\begin{figure}[t!]
\centerline{\includegraphics [scale = 0.5] {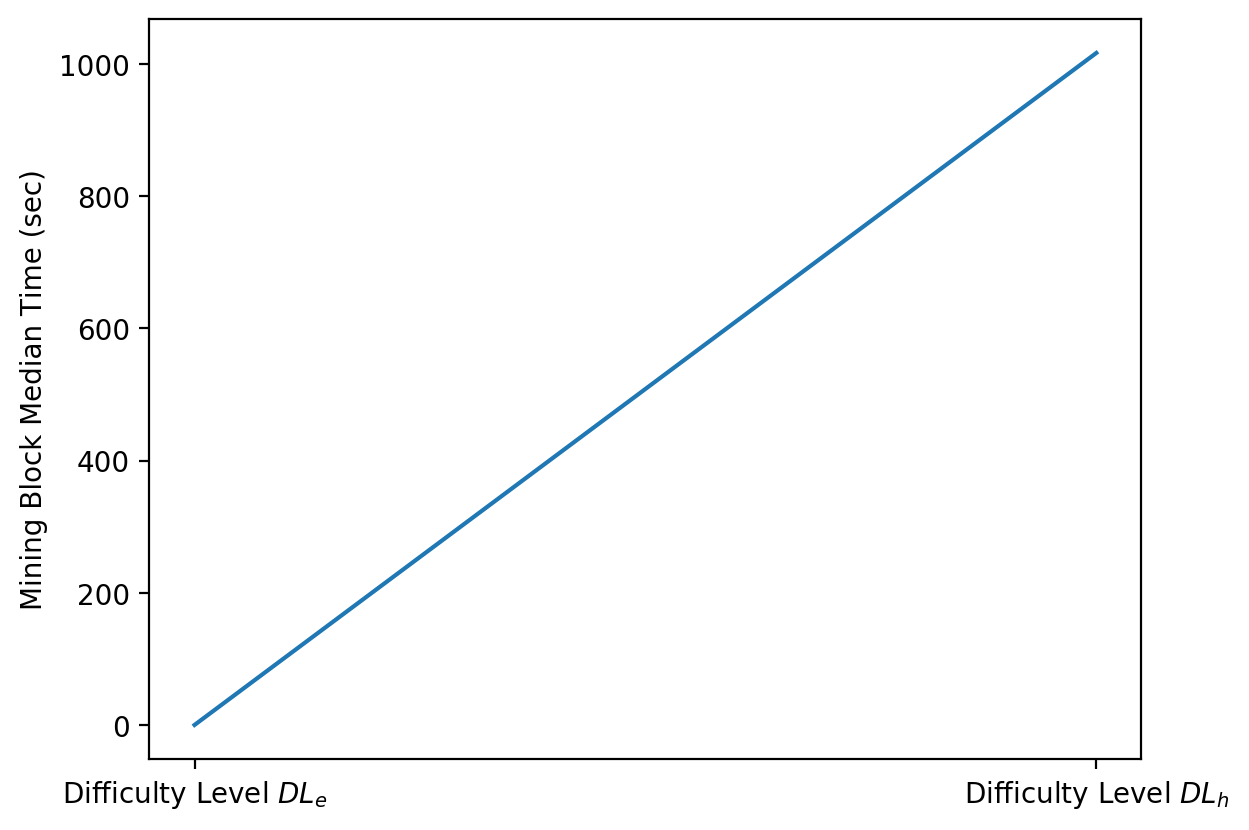}}
\vspace{-.15in}
\caption{\footnotesize Difficulty Level and Mining Time relation in $\ICT$.}
\label{DLeh}
\end{figure}
In our current evaluation setup, we mine $120$ blocks and import a mean of $200$ number of transactions to each of them.  Fig.~\ref{DLeh}, shows the median time spent to mine the block for different difficulty levels. Table~\ref{Table:table2}, represents the maximum, minimum, and average time required to mine a block based on different difficulty levels. The time spent to mine these blocks based on the W-Hash values completely defines the superiority of the dynamic approach to make the mining process easier for high-credit nodes and legally authorized entities. As the dPoW is based on randomly selecting the W-Hash value for the current block to make the anomaly behavior harder to conduct, the average time spent to mine the block based on two difficulty levels can be a reasonable metric to contrast the applicability of the system. $DL_e$ is far much smaller than $DL_h$, providing a much easier procedure for the high credit nodes to mine a block. While in Bitcoin~\cite{Bitcoin}, every $10$ minute, a new block can be mined, this process in the $\ICT$ framework is much easier and sooner for the high credit nodes. A high credit miner in this network can mine more than one block in each second and add the necessary transactions to the network in a timely fashion. In contrast, this process for the malicious nodes is much longer. Before conducting any attack in the network and any attempt to change the data of the previous blocks, other blocks will be added to the network via the honest nodes, and tampering with the signed distributed data in the network will become harder.
\begin{table}
\centering
\caption{\small Difficulty Level (DL), in the Proposed Dynamic PoW.}\label{Table:table2}
\begin{tabular}{| l | l | l | l | l | l |}
\hline
\textbf{DL} & \textbf{Size} & \textbf{Target Hash} & \textbf{Min. Time} & \textbf{Max. Time} & \textbf{Ave. Time} \\ \hline
$DL_e$ & 4 & SHA256[0:1] & 0.0018 & 0.9287 & 0.34501 \\ \hline
$DL_h$ & 16 & SHA256[0:4] & 3.823 & 4986.3987 & 1286.6703 \\ \hline
\end{tabular}
\end{table}

\vspace{.05in}
\noindent
\textit{A. Localization Performance}

\begin{figure}[t!]
\centerline{\includegraphics [scale = 0.4] {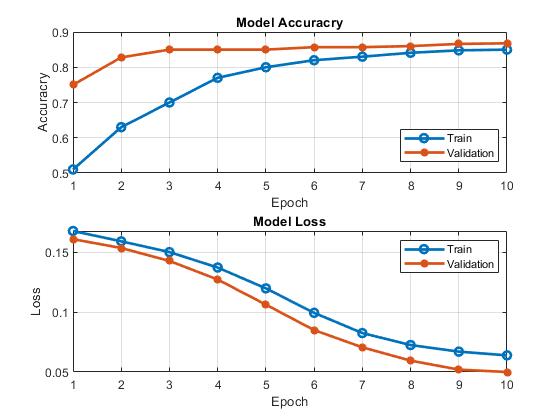}}
\vspace{-.15in}
\caption{\footnotesize Accuracy and loss of the proposed CNN-based AoA~scheme. \label{model_acc}}
\vspace{-.2in}
\end{figure}
\begin{figure}[t!]
\centerline{\includegraphics [scale = 0.4] {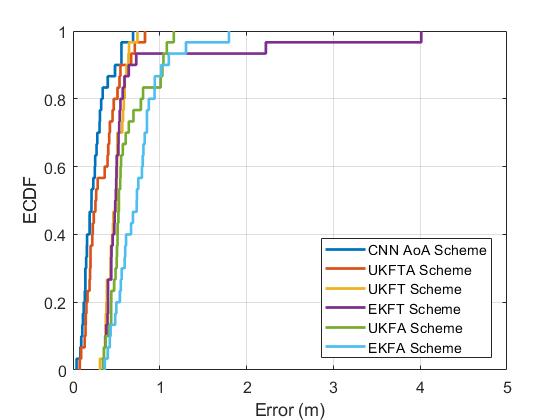}}
\vspace{-.15in}
\caption{\footnotesize Location error ECDF. \label{ECDF}}
\vspace{-.2in}
\end{figure}
Each user can be localized through a set of BLE beacons if the corresponding user is located in the receptive field of that BLE. The localization dataset is generated based on three different channel models: $(i)$ AWGN model, where $10 \leq \text{SNR} \leq 20$ dB; $(ii)$ Rayleigh fading channel, and; $(iii)$ Rayleigh fading channel affected by noise in a 3-D indoor environment. We also consider the destructive effect of the elevation angle on the special spectrum of the AoA measurements in this dataset. We use $76,545$ angle images labeled by their location for training. Fig.~\ref{Angle}(b) illustrates a typical angle image, containing $724$ sets of AoA measurements generated by the subspace-based algorithm.
The original angle image is of size $4 \times 181$, reshaped to be a square angle image of size $28 \times 28$ by zero padding. For the validation and test set, we use $15,309$ and $10,206$ angle images, respectively, which are all previously unseen and randomly chosen.

The accuracy and the loss of the proposed CNN-based AoA framework versus the number of epochs are represented in Fig.~\ref{model_acc}. As it can be seen from Fig.~\ref{model_acc}, the model accuracy increases and the loss decreases in each epoch, which means the model is well trained. Moreover, the accuracy of the proposed localization platform over the test set is $87\%$, which is a high location accuracy in the presence of noise, Rayleigh fading, and elevation angle. To illustrate the superiority of the proposed approach, we compare the Empirical Cumulative Distribution Function (ECDF) for location error. As it can be seen from Fig.~\ref{ECDF}, CNN-based AoA framework is compared with Extended Kalman Filter AoA (EKFA), EKF Time Difference of Arrival (TDoA) (EKFT), Unscented Kalman Filter AoA (UKFA), UKF ToA (UKFT)~\cite{Yousefi2015}, and UKF TDoA AoA (UKFTA) frameworks. According to the results, the location error of the CNN-based AoA framework is significantly lower than that of its counterparts. By leveraging the proposed CNN-based AoA localization framework implemented and embedded in our blockchain-based application, the proposed $\ICT$ framework can positively prevent the spreading of the COVID-19.

\begin{figure}[t!]
\centerline{\includegraphics [scale = 0.65] {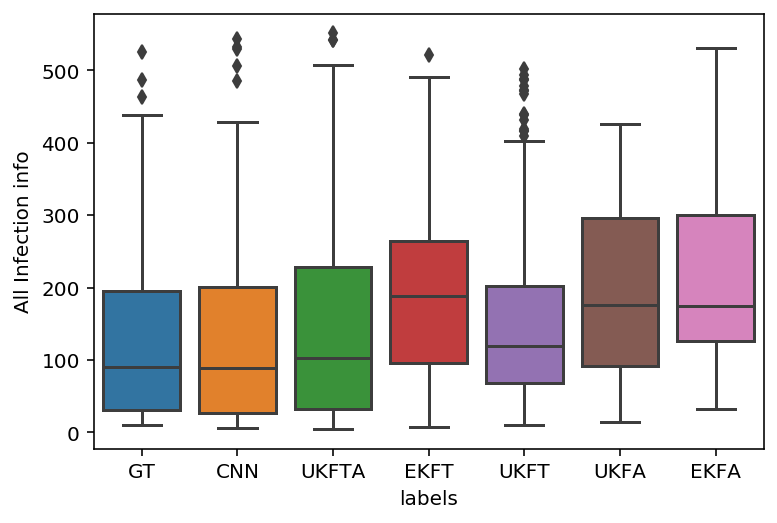}}
\vspace{-.15in}
\caption{\footnotesize Number of users infected in the test indoor environment because of one infection case when a different proximity recognition algorithm is used.}
\label{noInfection2}
\vspace{-.2in}
\end{figure}

\begin{figure}[t!]
\centerline{\includegraphics [scale = 0.6] {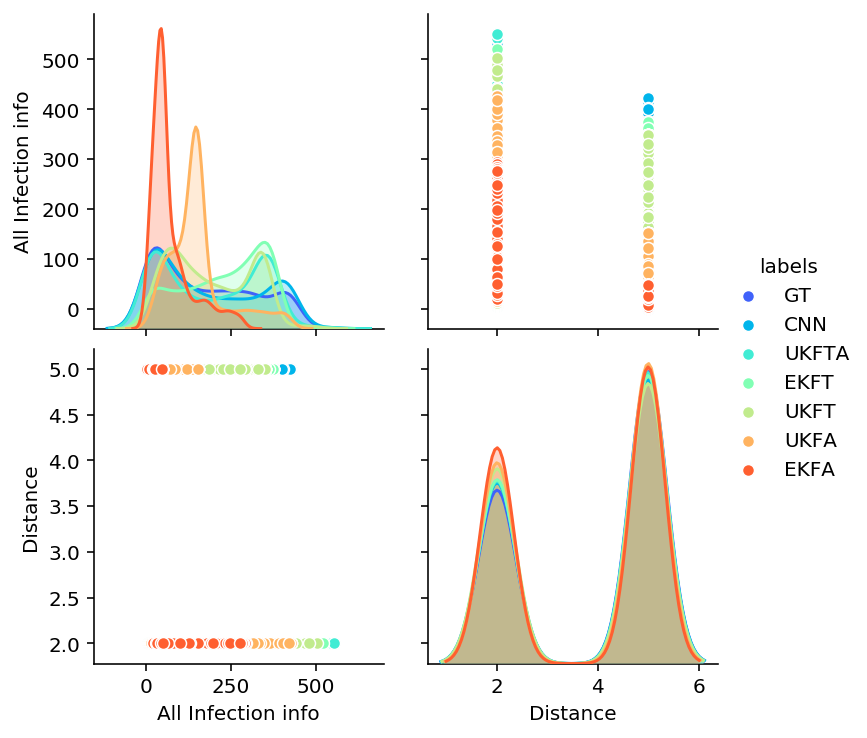}}
\vspace{-.15in}
\caption{\footnotesize Infection ratio based on two different social distance measures, i.e., 2 meters and 5 meters.}
\label{bothDistances}
\vspace{-.2in}
\end{figure}
\begin{table*}
\centering
\caption{\small Average number of interactions of infections out of $1,000$ users in indoor environment.}\label{Table:table4}
\begin{tabular}{| l | l | l | l | l | l | l |}
\hline
\textbf{Metric} & \textbf{CNN} & \textbf{UKFTA} & \textbf{EKFT} & \textbf{UKFT}& \textbf{UKFA} & \textbf{EKFA} \\ \hline
\textbf{Avg. No. Interactions} & \textbf{76,306.118} & 75,770.248 & 75,803.404 & 75,640.436 & 72,247.544 & 72,811.806 \\ \hline
\textbf{Avg. Gained Credit} & \textbf{57,111.127} & 49,716.447 & 43,178.33 & 46,985.936 & 2,995.165 & 26918.931 \\ \hline
\end{tabular}
\end{table*}
\begin{figure*}[t!]
\centering
\subfigure[]{\includegraphics[width=0.48\textwidth]{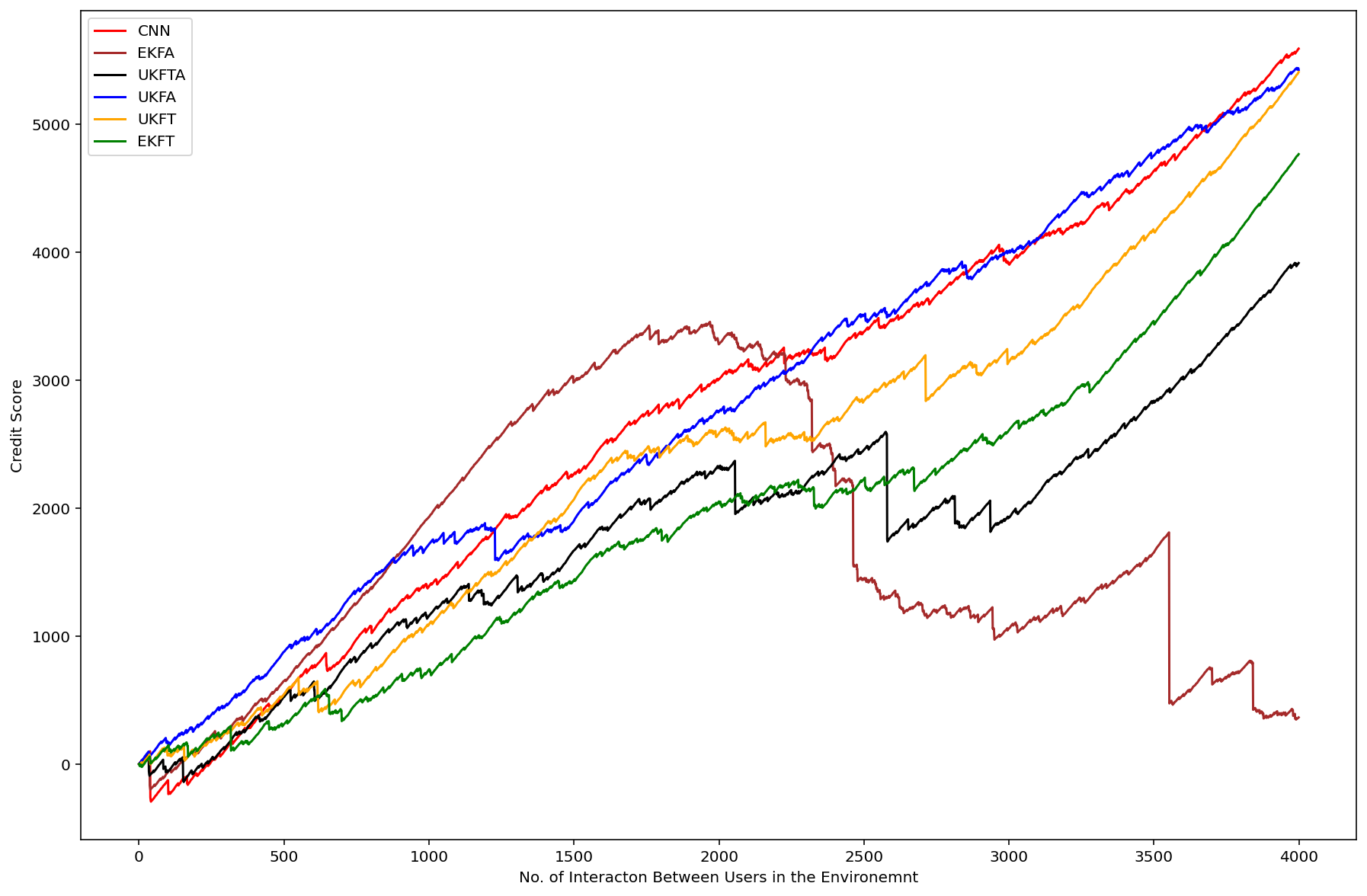}}
\subfigure[]{\includegraphics[width=0.48\textwidth]{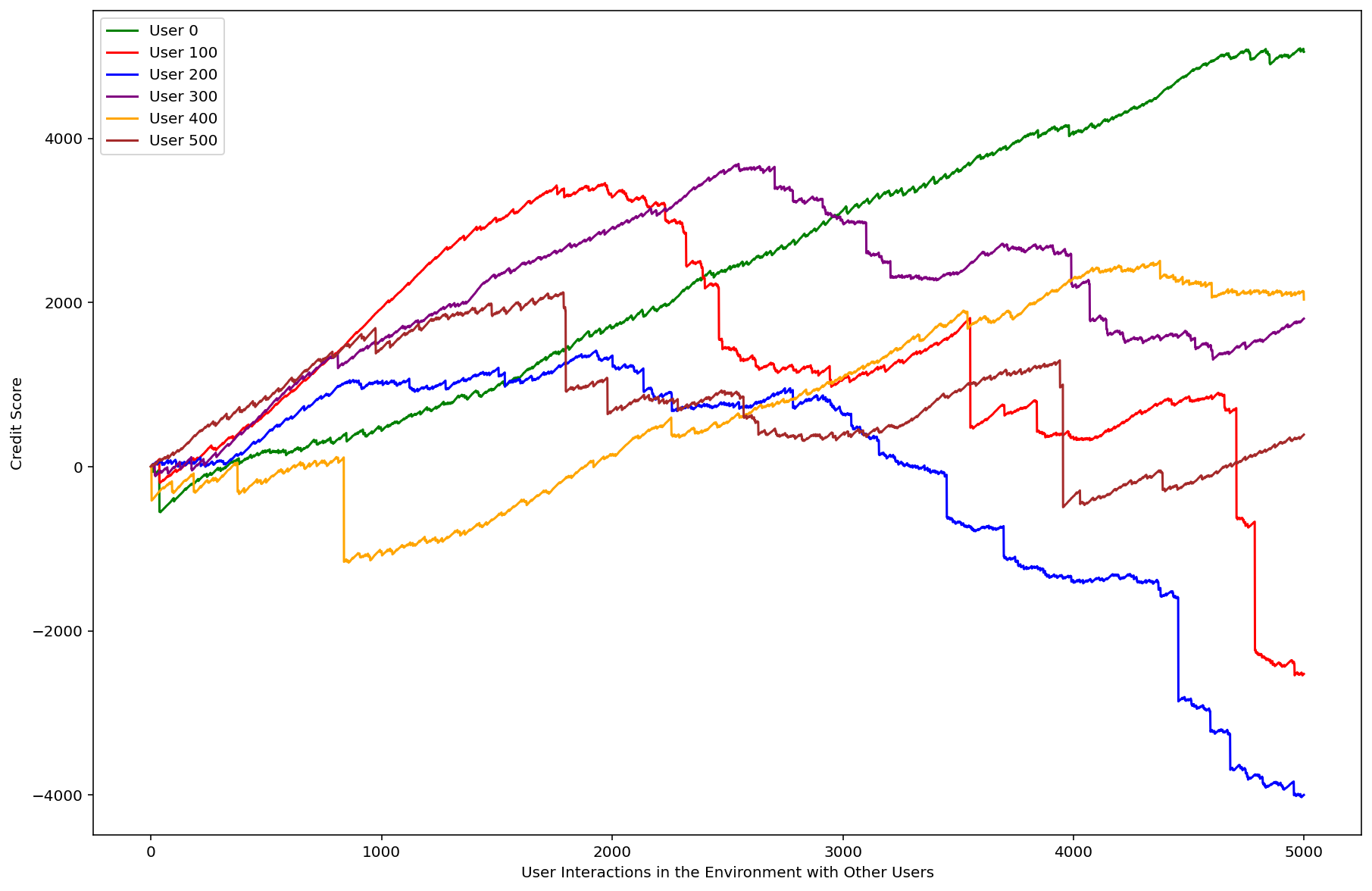}}
\vspace{-.15in}
\caption{Credit Score Gained by Users in the Network Considering Different Localization Algorithms. (a) Average Gained Credit Scores in First 4000 Interactions (b) Credit Scores Gained By Six Sample Users in CNN-Based Localization Approach.}
\label{credit12}
\vspace{-.25in}
\end{figure*}

\begin{figure}[t!]
\centerline{\includegraphics [scale = 0.35] {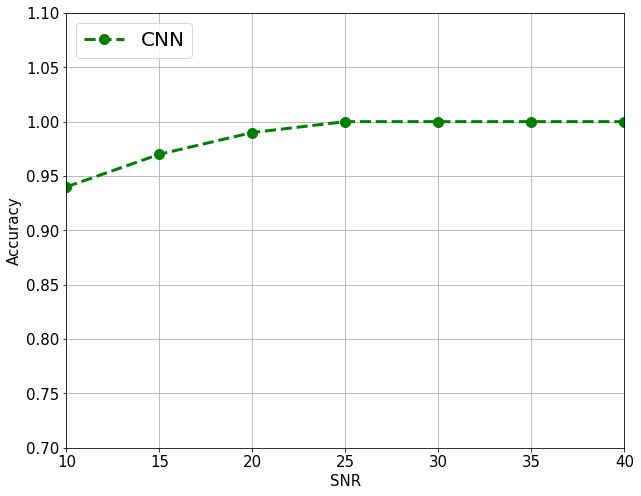}}
\vspace{-.15in}
\caption{\footnotesize Location accuracy versus different values of SNR (dB).}
\label{SNR}
\vspace{-.2in}
\end{figure}

To evaluate the feasibility of the trustworthy blockchain-enabled system in terms of the noise level, we have computed the location accuracy of the TB-ICT framework (after a specific epoch where the CNN model is well trained) versus different SNR values. Note that the common value of SNR in wireless networks is in the range of $15 \leq SNR \leq 40$ dB~\cite{Stack_Exchange}, where the SNR greater than $40$ dB and lower than $15$ dB are considered excellent and unreliable connections, respectively. Here, the localization dataset is affected by noise, which is modeled by AWGN with $10 \leq \text{SNR} \leq 20$ dB. Fig.~\ref{SNR} illustrates the impact of the SNR on the location accuracy. According to the results in Fig.~\ref{SNR}, increasing the SNR value leads to decreasing the location error. By considering the effect of the low SNR in the range of $10 \leq SNR \leq 20$ dB on the train dataset of the proposed TB-ICT framework, it can be observed from Fig.~\ref{SNR} that the TB-ICT framework is robust against noise.

\vspace{.025in}
\noindent
\textit{Infection Spreading Nature:}
Given the location of users during their movements and to illustrate the performance of the proposed $\ICT$ framework, first, we investigate how one infected user can infect other users in its close contact. Fig.~\ref{noInfection2} evaluates the disease spreading trend. According to the results, one infected user can infect at least $341$ users (out of $1000$) by considering $2$m as the required social distance criteria. Each of these users themselves can infect a large number of users in the environment (second cycle of infection). The contagiousness can be worsened by considering $5$ m as shown in Fig.~\ref{bothDistances}).

\vspace{.05in}
\noindent
\textit{B. Evaluation of the Credit-based Mechanism}

The credit-based approach can assist with management of the pandemic promoting users to act rationally. As stated previously, abnormal behaviors can be classified into two main categories, i.e., (i) Disease-Related Violations such as breaching the physical distancing rule and the rules related to the disease mostly bond to the proximity-based localization results, and (ii) Anomalous Behavior in Blockchain Network including different attacks or wrong claims, e.g., false claims of infection to the COVID-19 or violating the rule of sharing the last 14 days contact information in case of being diagnosed as infected.
In our experiments, $1,000$ users are moving in an indoor venue having interactions with each other. Table~\ref {Table:table4} describes these interactions and related credit points exchanged between users based on different algorithms. It can be seen that $\ICT$ has better results in gaining credits while the interactions are almost the same. In Fig.~\ref{credit12}(a), average credit scores gained in the first $4,000$ interactions between one random user in the network considering different localization algorithms is illustrated. Agents following the $\ICT$ approach are gaining more credit in the environment. This calculation of the credit is based on Eq.~\eqref{e2}, while the $\lambda^- = 12$ and $\lambda^+ = 2$ are considered as the credit calculation coefficients. Alternative coefficient values can be considered based on the pandemic situation and the community reactions. In Fig.~\ref{credit12}(b), credit scores of six sample users out of all $1,000$ users are shown.

\begin{figure}[t!]
\centering
\includegraphics[width=0.48\textwidth]{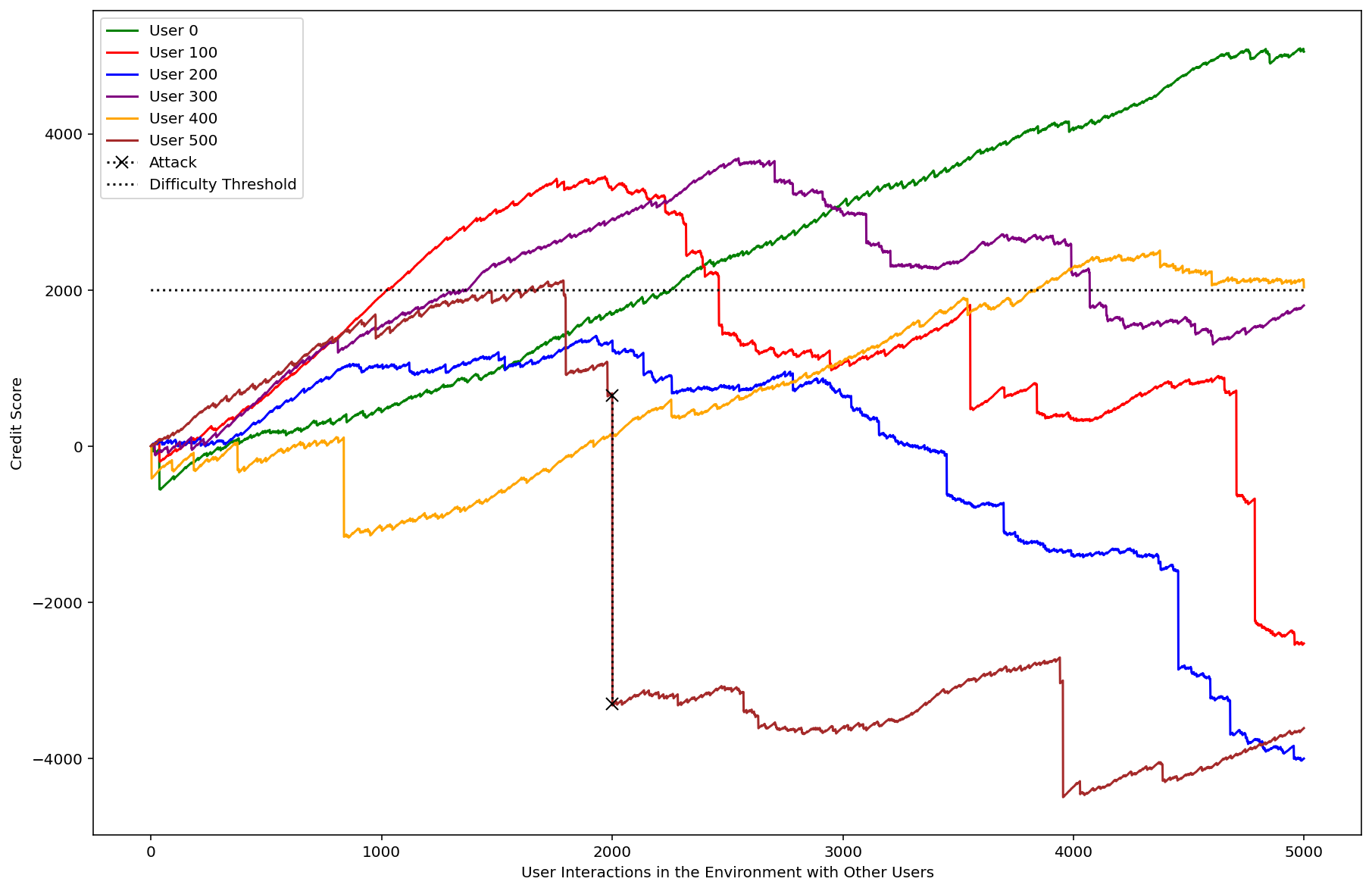}
\vspace{-.15in}
\caption{An attack scenario illustrating network's behaviour to punish the attacker (User 500).}
\label{attack12}
\vspace{-.25in}
\end{figure}

Any anomalous behavior will also be punished heavily in the $\ICT$ network. As Fig.~\ref{attack12} shows, in our experiment, each malicious activity, e.g., attacks in the network, will be punished with a high negative credit score. The nodes that are punished based on these attacks lose the chance to be an honest node in the network or act as a miner. Moreover, Fig.~\ref{attack12} represents the difficulty level ($\alpha_d$). Nodes with credit values above the threshold line can be considered as honest nodes and participate in the mining process. Difficulty level $1$ ($DL_e$) has the lowest difficulty with considering the first one significant digit of the block hash string. The first level of difficulty is applied to the legal, authorized nodes and the nodes in the network with a credit bigger than a threshold ($\alpha_d$). As can be seen, malicious behavior directly affects the credit values and can push the nodes' credits below the threshold line. Both nodes $0$ and $500$ experience this situation and losing their chance to be above the threshold line.

\vspace{.05in}
\noindent
\textit{C. Security Analysis}

As final step, we focus on the security and trustworthiness of the proposed framework in different aspects as below:

\noindent
\textit{Removing Centralized Trust, Single Point of Failure and Sybil attack:}
The $\ICT$ platform is a permissionless network without any dependency on centralized trust. Any central dependency increases the risk of data leakage and makes the system susceptible to the single point of failure issue. The $\ICT$ is a distributed ledger that inherently has several full nodes having the replicated copy of the whole system. Consequently, the system is resilient to the failure of some of the nodes in the network. Different characteristics of the proposed $\ICT$ solution, specifically the dynamic W-Hash proof of credit platform, coupled with inherited features of a blockchain network (being resilient to tampering with the mined and signed information) can be leveraged to guarantee the refusal of the unauthorized and malicious nodes.

\noindent
\textit{Guaranteeing the Validity of Shared Data:} In a pandemic crisis such as COVID-91, any fake information or false claims can produce a panic in public, therefore, accuracy and correctness of the personalized claims are of vital importance. The proposed dPoC algorithm will prevent publishing any false claims in the network. False or abnormal claims from any node in the network can have significant negative effects on the credit score and pushing the malicious nodes to participate under harder difficulty constraints. The higher level of difficulty, inherently, will prevent a potential attack or abnormal behavior in the network by maximizing the cost of malicious~activities.

\noindent
\textit{Secure Data Management:} Only the unique IDs of the nodes will be shared in the network. When a claim is transmitted in the case of a positive infection result, only the contacts' IDs will be shared in the network. All users' personal information is kept locally on their devices, without the need to be shared. Additionally, data encryption is implemented exploiting ECC with ECDSA to guarantee data~security.

\noindent
\textit{Computational Complexity:} Considering Eq.~\eqref{e7} and Algorithm~\ref{Alg1}, hash of the current block ($B_c$) (if the block is not already mined) is the hash of the concatenation of the current block and $n_{wh}-1$ of the previous blocks with the nonce value. $n_{wh}$ is the current $W-Hash$ value, which is defined and shared between the high-credit and legally authorized nodes based on their credit. To mine the block, miners compete to solve a mathematical puzzle and find the target hash string by adding the nonce in each step via a trial and error procedure. The target hash string in the $\ICT$ platform has $64$ digits in hexadecimal base. A compressed representation (called ``Bits") of the target hash string hex value, e.g., $0x1b0404cb$ is stored in the blocks. The hexadecimal value is calculated based on this packed form. For example $0000$ $0000$ $2FF4$ $04CB$ $0000$ $0000$ $0000$ $0000$ $0000$ $0000$ $0000$ $0000$ $0000$ $0000$ $0000$ $0000$, hexadecimal string of length $64$ and the hash value should be less than this value to mine a block. The leading zeros (eight leading zeros $0000$ $0000$) in the target hash string represent the difficulty level. As can be seen in Table~\ref{Table:table1}, for the nodes with higher credits and the legally authorized nodes, the first significant digit ($1$ digit) of the block hash string is considered the leading zeros. This value will be four significant digits of the block target hash string for the nodes with bad credit scores. To be more precise, lower target values will lead to the greater difficulty of block mining.

Assuming the length of target hash string is $b$ bits, for a $W-Hash$ value of size $n_{wh}$, computational complexity is of $O(n_{wh} \times t_{TH})$. Term $t_{TH}$ is the required time to generate the target hash string in the search space of $2^b$. Therefore, computational complexity is of $O(n_{wh} \times 2^b)$. The malicious node is now forced to find the hashes for all the possible values of the $W-Hash$ sizes, which directly increases its computational complexity and, consequently, the attack's cost. The complexity in this scenario is increased as follows
\begin{eqnarray}\label{e10}
Complexity = O(n_{wh} \times (n_{wh} \times t_{TH})) = O(n_{wh}^2 \times 2^b)
\end{eqnarray}
This attribute also makes other possible attacks such as double-spending and spamming infeasible. Without the W-Hash, the complexity of the network to conduct an attack would be as low as $O(t_{TH})$, while this complexity, after applying the hash window, will be $O(n^2 \times t_{TH} )$, which requires $n \times n$ more test and operation to find the block's target hash string. Consequently, the security of the network is significantly improved. On the other hand, based on the credit-based platform, any malicious attempts to even break this difficulty level will have an irreversible and high credit punishment, causing even higher computational costs to the faulty nodes.

\noindent
\textit{Consistency of $\ICT$ Blockchain:} In classical blockchain networks, consistency is provided via all miners' agreement on the longest chain, while synchronous updates about the latest block may be unavailable for all the miners at a specific time. More precisely, some miner nodes may receive the latest block, while some others not. The $\ICT$ framework can offer a perfect solution for consistency based on the high credit nodes in the network. Keeping track of the credit value of the nodes in the network enables the system to recognize the minimum required number of honest mining nodes that can acknowledge the current length of the chain and guarantee the agreed chain will not be removed or lost from the current chain of the blockchain.

\section{Conclusion}  \label{sec:con}
In this paper, we proposed a trustworthy blockchain-enabled system for the indoor COVID-19 contact tracing model, which securely identifies contact history of COVID-19 patients in indoor environments. To improve the accuracy of the CT model, we deployed an AI-based localization approach based on BLE sensor measurements. More precisely, a CNN-based AoA estimation framework is applied on the subspace-based angle estimation in a 3-D indoor environment to avoid the need for analytical modeling of noise, path-loss, and multi-path effects. The simulation results showed that the proposed $\ICT$ prevents the COVID-19 from spreading by the implementation of a highly accurate contact tracing model while improving the users' privacy and security. The rationale for deploying BLE sensors during the current pandemic time is that they are more ubiquitous in comparison to other technologies such as Ultra-Wideband (UWB). With the emphasis on higher achievable accuracy, a fruitful direction for future research is the use of hybrid UWB/BLE technologies.

\bibliographystyle{IEEEbib}

\bibliographystyle{IEEEtran}
\bibliography{keylatex}
\begin{IEEEbiography}[{\includegraphics[width=1in,height=1.25in,clip,keepaspectratio]{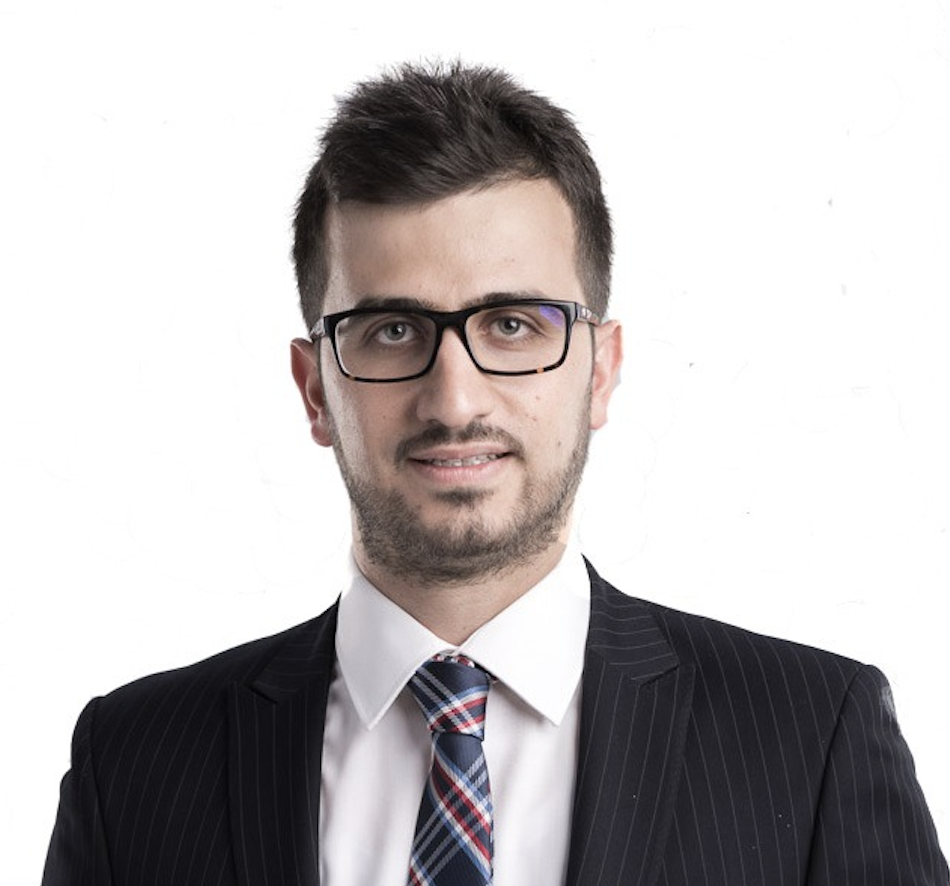}}]{Mohammad Salimibeni}, received the B.Sc. degree in Electrical Engineering from the Isfahan University of Technology (IUT), in 2011 and the M.Sc. degree in Electrical Engineering from the same university (IUT) in 2014. He is currently pursuing the Ph.D. degree at Concordia University, Montreal, Canada. As a member of I-SIP Lab, His research interests include signal processing, Reinforcement Learning and distributed IoT-based networks with a particular interest in distributed ledger technology and Blockchain.
\end{IEEEbiography}

\begin{IEEEbiography}[{\includegraphics[width=1in,height=1.25in,clip,keepaspectratio]{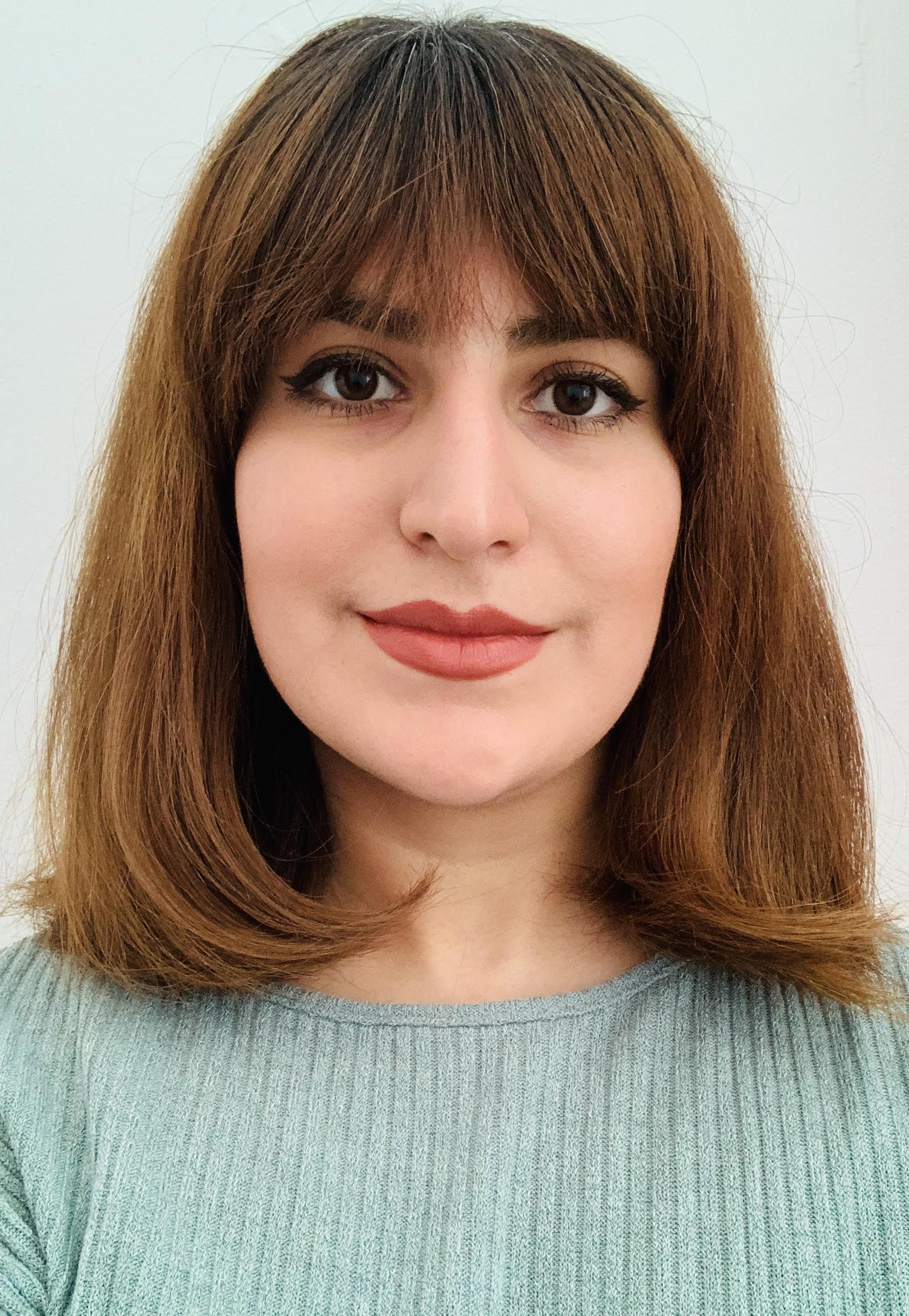}}]
{Zohreh Hajiakhondi-Meybodi} received the B.Sc. degree in Communication Engineering from Yazd University, Yazd, Iran and the M.Sc. degree in Communication Systems Engineering (with the highest honor) from Yazd University, Yazd, Iran in 2013 and 2017, respectively. She is a Ph.D. degree candidate at Electrical and Computer Engineering (ECE), Concordia University, Montreal, Canada. Since 2019, she has been an active member of I-SIP Lab at Concordia University. Her research interests include general areas of wireless communication networks with a particular emphasis on Femtocaching, Internet of Things (IoT), Indoor Localization, Optimization Algorithms, and Multimedia Wireless Sensor Networks (WMSN).
\end{IEEEbiography}
\begin{IEEEbiography}[{\includegraphics[width=1in,height=1.25in,clip,keepaspectratio]{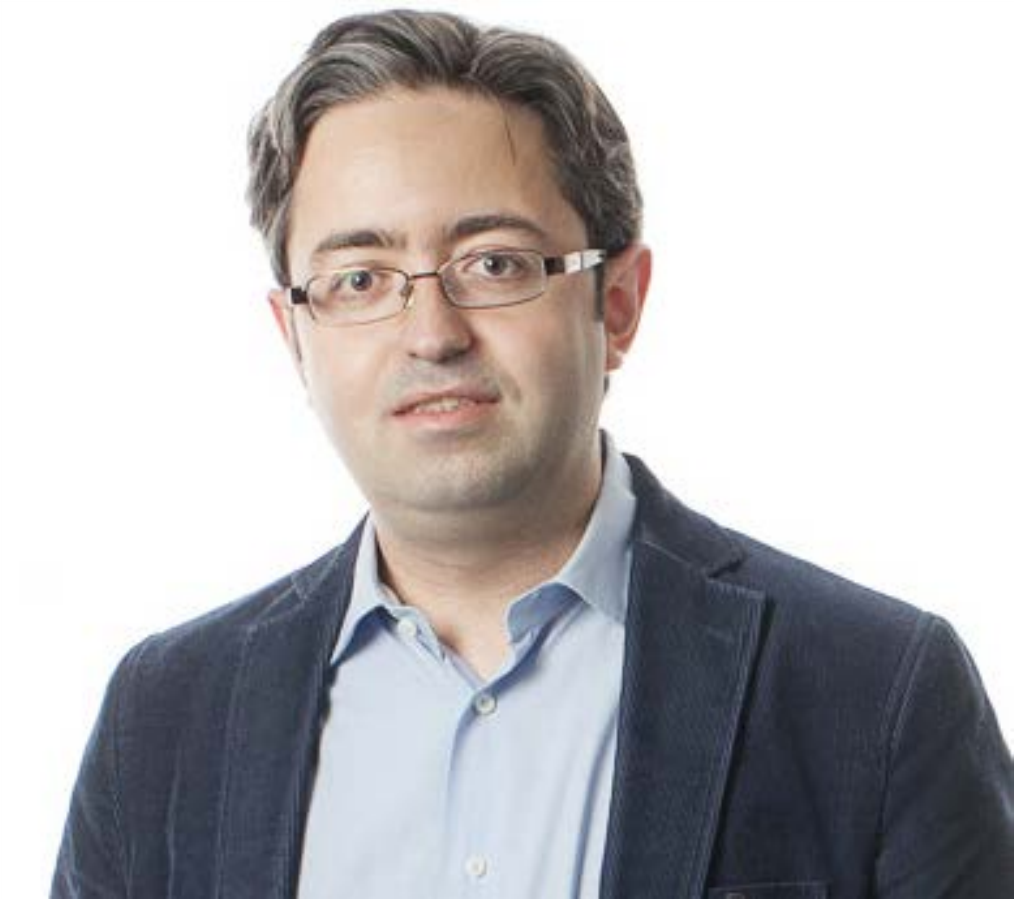}}]
{Arash Mohammadi}  (S'08-M'14-SM'17) is currently an Associate Professor with Concordia Institute for Information Systems Engineering, Concordia University, Montreal, QC, Canada. Prior to joining Concordia University and for 2 years, he was a Postdoctoral Fellow with the Department of Electrical and Computer Engineering, University of Toronto, Toronto, ON, Canada. Dr. Mohammadi is a registered professional engineer in Ontario. He is Director-Membership Developments of  IEEE Signal Processing Society (SPS); General Co-Chair of ``2021 IEEE International Conference on Autonomous Systems (ICAS),'' and; Guest Editor for IEEE Signal Processing Magazine (SPM) Special Issue on ``Signal Processing for Neurorehabilitation and Assistive Technologies''. He is also currently serving as Associate Editor on the editorial board of IEEE Signal Processing Letters. He was Co-Chair of ``Symposium on Advanced Bio-Signal Processing and Machine Learning for Assistive and Neuro-Rehabilitation Systems'' as part of 2019 IEEE GlobalSIP, and ``Symposium on Advanced Bio-Signal Processing and Machine Learning for Medical Cyber-Physical Systems,'' as a part of IEEE GlobalSIP'18; The Organizing Chair of 2018 IEEE Signal Processing Society Video and Image Processing (VIP) Cup, and the Lead Guest  Editor for IEEE Transactions on Signal \& Information Processing over Networks Special Issue on ``Distributed Signal Processing for Security and Privacy in Networked Cyber-Physical Systems''. He is recipient of several distinguishing awards including the Eshrat Arjomandi Award for outstanding Ph.D. dissertation from Electrical Engineering and Computer Science Department, York University, in 2013; Concordia President's Excellence in Teaching Award in 2018, and; 2019 Gina Cody School of Engineering and Computer Science's Research and Teaching awards .
\end{IEEEbiography}

\begin{IEEEbiography}[{\includegraphics[width=1in,height=1.25in,clip,keepaspectratio]{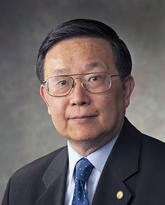}}]{Yingxu Wang} is professor of cognitive informatics, brain science, software science, and denotational mathematics. He is the founding President of International Institute of Cognitive Informatics and Cognitive Computing (ICIC). He is a Fellow of ICIC, a Fellow of WIF (UK), a P.Eng. of Canada, and a Senior Member of IEEE and ACM. He has been visiting professor (on sabbatical leaves) at Oxford University (1995), Stanford University (2008 | 2016), UC Berkeley (2008), and MIT (2012), respectively. He has been a full professor since 1994. He is the founder and steering committee chair of the annual IEEE International Conference on Cognitive Informatics and Cognitive Computing (ICCI*CC) since 2002. He is founding Editor-in-Chiefs of International Journal of Cognitive Informatics \& Natural Intelligence; International Journal of Software Science \& Computational Intelligence; Journal of Advanced Mathematics \& Applications; and Journal of Mathematical \& Computational Methods, as well as Associate Editor of IEEE Trans. on SMC - Systems. He has served as chair or co-chair of 19 IEEE or other int’l conferences and a member of IEEE Selection Committee for Senior Members in 2011.
Dr. Wang’s publications have been cited for 15,000+ times according to Google Scholar. According to Research Gate statistics, his research profile has reached top 2.5 per cent worldwide, top one to 10 (timely vary in the range) at the University of Calgary, and the most read work in neural networks. He is the recipient of dozens international awards on academic leadership, outstanding contributions, best papers, and teaching in the last three decades.
\end{IEEEbiography}

\end{document}